\documentclass[fleqn,usenatbib]{mnras}

\usepackage{newtxtext,newtxmath}

\usepackage[T1]{fontenc}

\DeclareRobustCommand{\VAN}[3]{#2}
\let\VANthebibliography\thebibliography
\def\thebibliography{\DeclareRobustCommand{\VAN}[3]{##3}\VANthebibliography}

\usepackage{graphicx}	
\usepackage{amsmath}	
\usepackage{amssymb}	

\title[First Order Resonance Widths]{On the Divergence of First Order Resonance Widths at Low Eccentricities}

\author[Renu Malhotra and Nan Zhang]{
Renu Malhotra,$^{1}$\thanks{E-mail: renu@lpl.arizona.edu}
Nan Zhang$^{2}$
\\
$^{1}$Lunar and Planetary Laboratory, The University of Arizona, Tucson, AZ 85721, USA\\
$^{2}$School of Aerospace Engineering, Tsinghua University, Beijing, 100084, China
}

\date{Accepted XXX. Received YYY; in original form ZZZ}

\pubyear{2020}

\begin{document}
\label{firstpage}
\pagerange{\pageref{firstpage}--\pageref{lastpage}}
\maketitle

\begin{abstract}
Orbital resonances play an important role in the dynamics of planetary systems.  Classical theoretical analyses found in textbooks report that libration widths of first order mean motion resonances diverge for nearly circular orbits. Here we examine the nature of this divergence with a non-perturbative analysis of a few first order resonances interior to a Jupiter-mass planet. We show that a first order resonance has two branches, the pericentric and the apocentric resonance zone. As the eccentricity approaches zero, the centers of these zones diverge away from the nominal resonance location but their widths shrink. We also report a novel finding of "bridges" between adjacent first order resonances: at low eccentricities, the apocentric libration zone of a first order resonance smoothly connects with the pericentric libration zone of the neighboring first order resonance. These bridges may facilitate resonant migration across large radial distances in planetary systems, entirely in the low eccentricity regime.
\end{abstract}




\section{Introduction}

When the orbital periods of two planets orbiting a central host star are close to the ratio of small integers, their mutual gravitational effects are enhanced and the orbital dynamics becomes quite complicated. This is the case even for the simplest model of the planar circular restricted three body problem (PCRTBP) in which one of the two planets is massless and is restricted to move in the orbital plane of the massive planet which itself moves in a fixed circular orbit. Although highly simplified, this model has many applications in planetary dynamics, such as the effects of a planet on minor planets (Jupiter on the asteroid belt, Neptune on the Kuiper belt), planetary moons on planetary ring particles, and the perturbations of exo-planets on debris disks. This simple model also provides the foundation for analyzing planet-planet interactions in the general case of multi-planet systems. With many decades of studies, we now understand that a mean motion resonance zone has non-zero measure in phase space, it contains stable and unstable equilibrium points (corresponding to periodic orbit solutions) and a domain of regular quasi-periodic orbits librating about the stable equilibrium (the resonance libration zone) bounded by a separatrix that passes through the unstable equilibrium point; in some cases, the separatrix dissolves into a chaotic layer in phase space. Thus, a mean motion resonance is a source of both stability and instability.
A central question is: what is the extent of the resonance zone which supports stable librations of resonant orbits? The answer is not easily won, in part because what at first sight appears to be a one-parameter problem (i.e., the period ratio) is actually a problem of at least two degrees of freedom, and in part because neighboring resonances interact with each other. Consequently, an accurate answer has been surprisingly elusive. 

In the regime of very low orbital eccentricities, we have been puzzled about the ambiguities of the resonance widths of first order mean motion resonances (MMRs) reported in many publications. To illustrate, we cite a current textbook on planetary dynamics \citep{Murray:1999SSD} which gives an analytical estimate based on perturbation theory that the resonance zone widths of Jupiter's first order MMRs in the asteroid belt diverge to infinity for circular orbits (see their Eq.~8.76 and their Fig.~8.7). \cite{Winter:1997} used a non-perturbative numerical approach (with Poincar\'e sections) to measure resonance widths, and they reported some discrepancies with the analytical perturbation theory estimates. On the same topic, the textbook by \cite{Morbidelli:2002Book} describes that the stable resonance center of Jupiter's 2/1 MMR diverges only on side of the nominal resonance location as $e$ approaches zero, and that a resonance separatrix vanishes for $e\lesssim0.2$ making the resonance width undefined for smaller eccentricities (see his Fig.~9.11.)  There are many other published studies exploring this topic with various levels of approximation and in various contexts; for recent examples we refer the reader to \cite{Mardling:2013}, \cite{Deck:2013}, \cite{Hadden:2018}, \cite{Namouni:2018}, and references therein.

In the work presented here we revisit the question of the sizes of the first order resonance zones as $e\longrightarrow0$. We adopt a non-perturbative approach based on computing Poincar\'e sections of the PCRTBP, as in \cite{Winter:1997}. Our implementation of this method is different than that of \cite{Winter:1997} who followed the lead of \cite{Henon:1966} on the choice of the Poincar\'e sections by recording the test particle's state vector at every successive conjunction with Jupiter. Many subsequent works on the  PCRTBP have adopted the same condition for the surface-of-section. However this is not a unique choice. In our implementation, we record the test particle's state vector at every successive perihelion passage. This choice is physically motivated so as to trace the behavior (libratory or not) of the test particle's perihelion longitude relative to Jupiter's position. This small but important change yields a more direct visualization and physical interpretation of the resonance libration zones in the Poincar\'e sections. \cite{Wang:2017} used this choice of the Poincar\'e sections to measure the widths of the 3/2 and 2/1 interior MMRs for a range of perturber masses and test particle eccentricities, $0.05 \le e \le 0.99$, but did not examine the regime of very low eccentricities. Here we investigate the very low eccentricity regime of Jupiter's interior 2/1, 3/2 and 4/3 MMRs. This yields a few new insights and more clarity on first order MMR widths at low eccentricities.

\section{Methodology}\label{s:methodology}

We follow the method described in \cite{Wang:2017} to compute the Poincar\'e sections, and from these we measure the resonance widths; we briefly describe this method here.  We adopt the circular planar restricted three body model of the Sun, Jupiter and test particle (the latter representing an asteroid).  In this approximation, all the bodies move in a common plane, Jupiter revolves around the Sun in a circular orbit, and the test particle revolves in an (osculating) elliptical heliocentric orbit. The masses of Sun and Jupiter are denoted by $m_1$  and $m_2$, respectively. The fractional mass ratio of Jupiter is very small, $\mu=m_2/(m_1+m_2)= 9.53 \times 10^{-4}$. We adopt the natural units for this model: the unit of length is the constant orbital separation of $m_1$ and $m_2$, the unit of time is their orbital period divided by $2\pi$, and the unit of mass is $m_1+m_2$; with these units the constant of gravitation is unity, and the orbital angular velocity of $m_1$ and $m_2$ about their common center of mass is also unity.
Then, in a rotating reference frame, of constant unit angular velocity and origin at the barycenter of the two primaries, both $m_1$ and $m_2$ remain at fixed positions, $ (-\mu,0) $ and $ (1-\mu,0) $, respectively, and we denote with ($x, y$)  the position of the test particle. The distances between the test particle and the two primaries can be written as
\begin{equation}\label{r1r2}
 r_1 =\left[ (x+\mu)^2+y^2 \right]^{1/2}\qquad \hbox{and}\qquad
 r_2 =\left[ (x-1+\mu)^2+y^2 \right]^{1/2},
\end{equation}	
and the equations of motion of the test particle can be written as,
\begin{eqnarray}\label{xdotydot}
\begin{cases}
\ddot{x}&=2\dot{y}+x-{{(1-\mu)(x+\mu)}\over {{r_1}^3}}-{{\mu(x-1+\mu)}\over{{r_2}^3}},
\\
\ddot{y}&=-2\dot{x}+y-{{(1-\mu)y}\over{{r_1}^3}}-{{\mu y}\over{{r_2}^3}}.
\end{cases}
\end{eqnarray}
where "$\cdot$" and "$\cdot\cdot$" represent the first and second derivative with respect to time.
These equations admit a conserved quantity, the Jacobi constant, $C_J$, given by,
\begin{equation}\label{jacobi}
C_J=x^2+y^2-\dot{x}^2-\dot{y}^2+\frac{2(1-\mu)}{r_1}+\frac{2\mu}{r_2},
\end{equation}
which can also be expressed approximately in terms of the Keplerian orbital elements, $ a $ and $ e $,  the semi-major axis and eccentricity, respectively, of the particle's heliocentric osculating orbit,
\begin{equation}\label{jacobiae}
C_J = \frac{1-\mu}{a}+2\sqrt{(1-\mu)a(1-e^2)}+O(\mu).
\end{equation}

\begin{figure}
 \centering
\hspace{-5pt} \includegraphics[width=100mm]{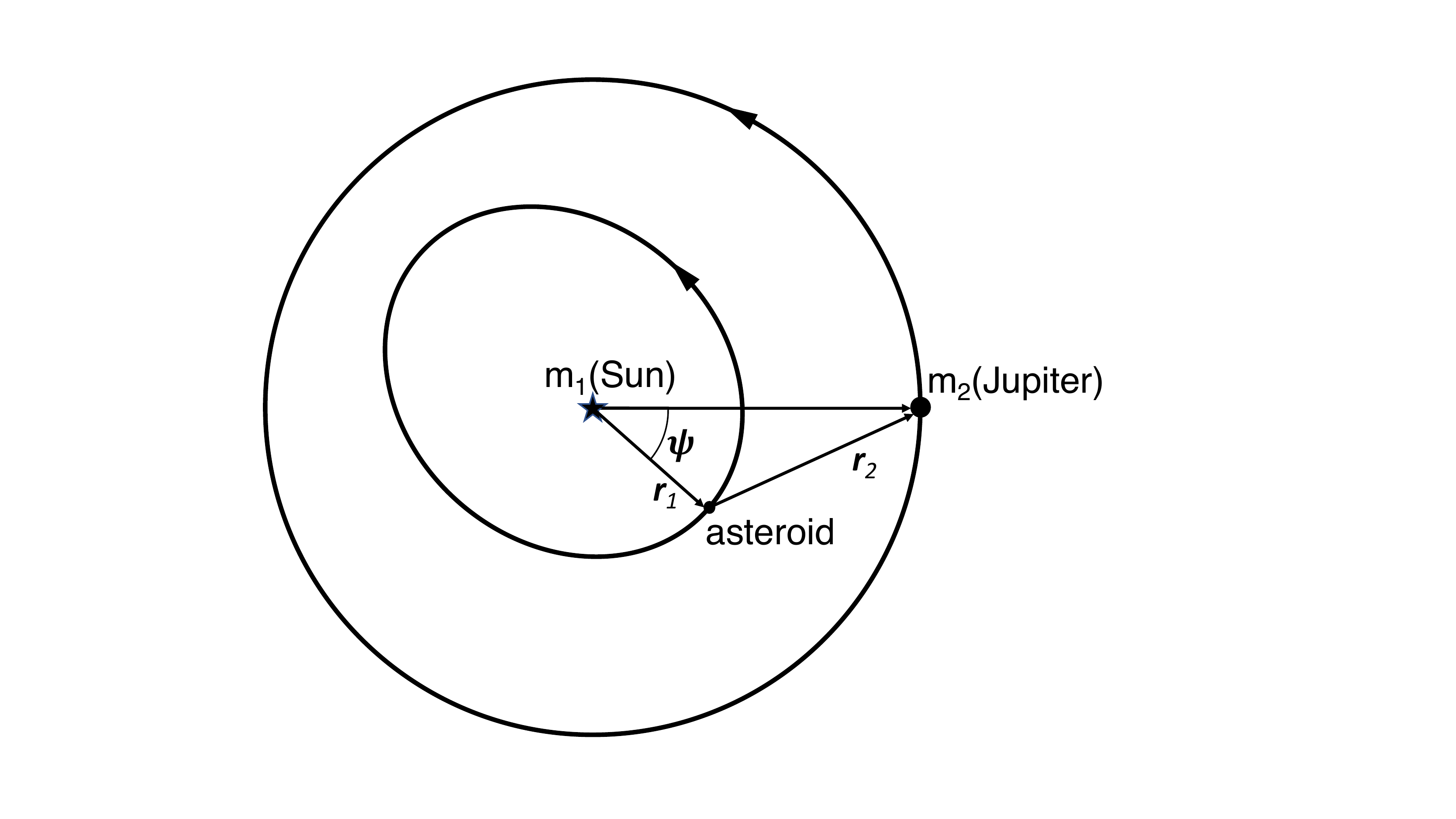}
 \caption{A schematic diagram to illustrate the definition of $\psi$, the angular separation of Jupiter from the test particle when the latter is at pericenter. }
 \label{f:f1}
\end{figure}

The test particle moves in a four dimensional phase space $(x,y,\dot{x},\dot{y})$.  
To conveniently and meaningfully visualize in a two dimensional plane the behavior of particles of different initial conditions but having the same Jacobi constant, we use the Poincar\'e surface-of-section technique, as follows. Starting from specified initial conditions, the equations of motion, Eq.~\ref{xdotydot}, are numerically integrated with the adaptive step size 7th order Runge-Kutta method~\citep{Fehlberg:1968} to obtain the continuous track of the test particle for several thousand Jupiter orbital periods. The relative and absolute error tolerances are controlled to be lower than $10^{-12}$.  
To obtain the Poincar\'e sections, we record the state vector, $(x,y,\dot{x},\dot{y})$, of the particle at every perihelion passage. 
We then transform these state vectors into osculating heliocentric orbital elements, semi-major axis $a$, eccentricity $e$, and the angle $\psi$ which measures the angular separation of Jupiter from the test particle when the particle is at perihelion; the definition of $\psi$ is illustrated in Figure~\ref{f:f1}. 

For future reference, we introduce the usual "critical resonant angle" defined by
\begin{equation}\label{e:phi}
\phi = (p+1)\lambda' - p\lambda-\varpi,
\end{equation}
where $\lambda,\lambda'$ are the mean longitudes of the particle and of Jupiter, respectively, and $\varpi$ is the particle's longitude of perihelion. For every point on the Poincar\'e section, the particle is located at its perihelion (that is, $\lambda=\varpi$), therefore $\phi=(p+1)(\lambda'-\varpi)$. Consequently, we have the following relationship between $\phi$ and $\psi$:
\begin{equation}
\phi = (p+1)\psi.
\label{e:phi-psi}\end{equation}

\begin{figure*}
\centering
   \begin{tabular}{c c c}
   \includegraphics[width=2.5in]{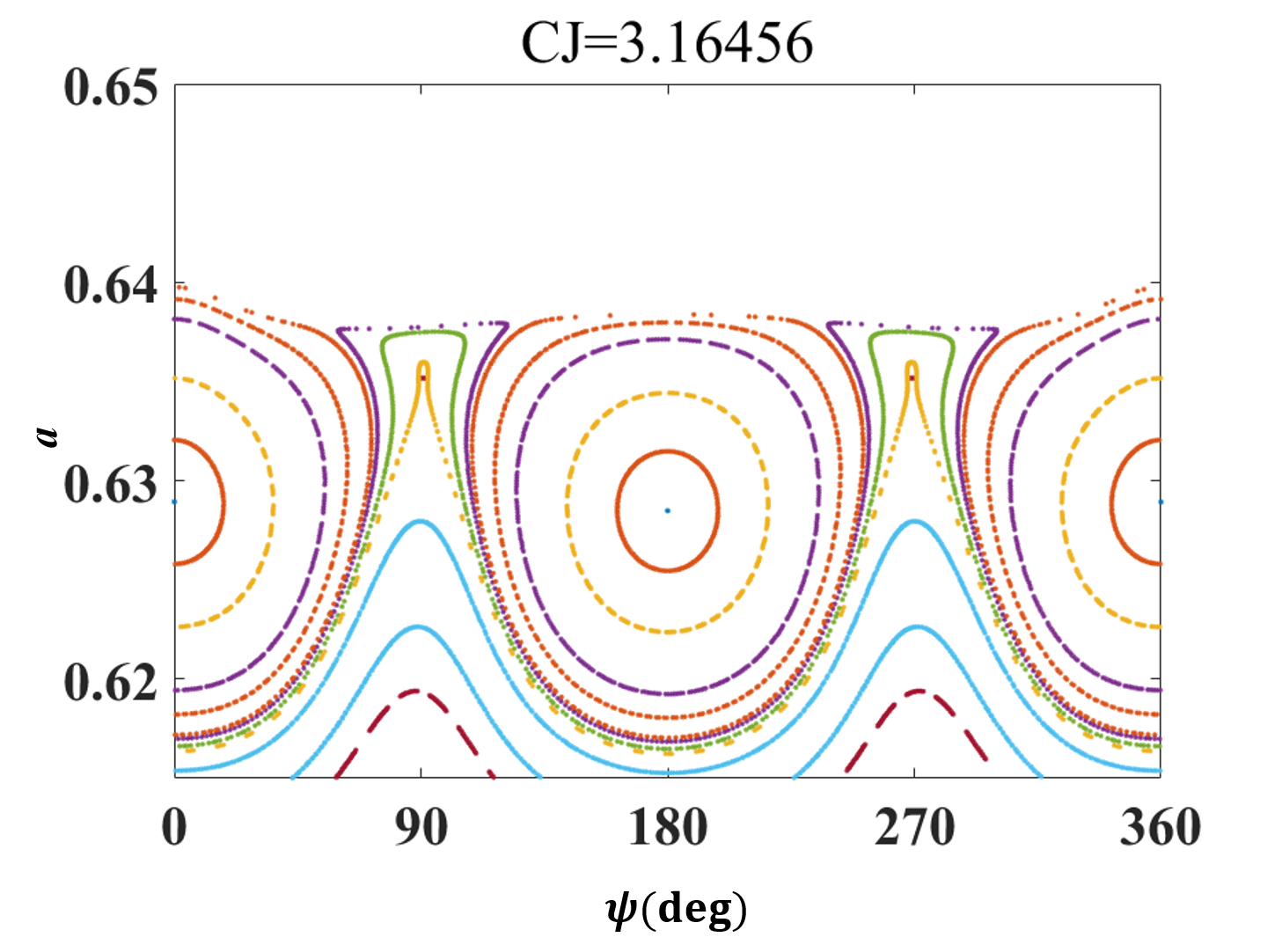} &
   \includegraphics[width=2in]{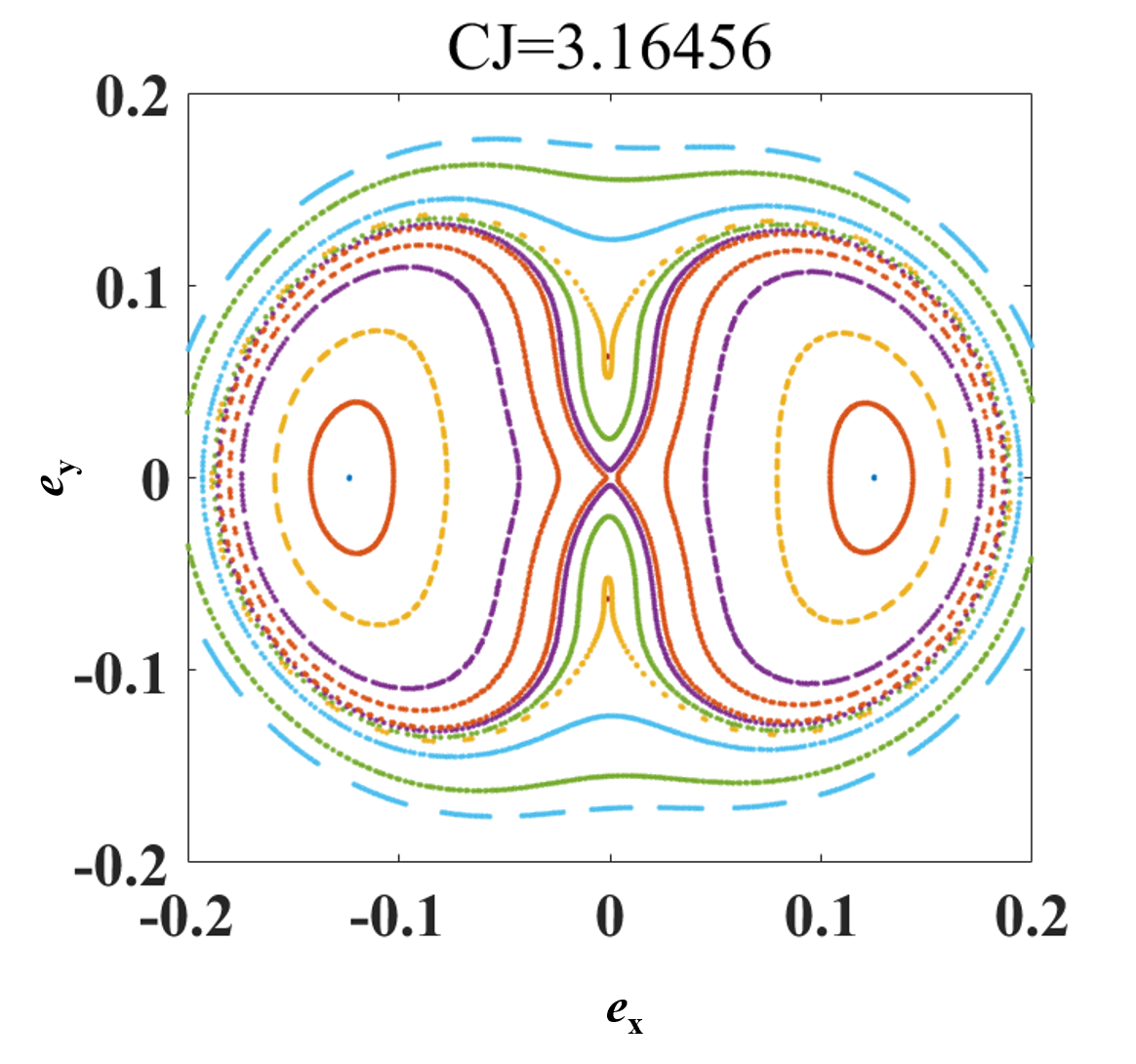} &
   \includegraphics[width=2in]{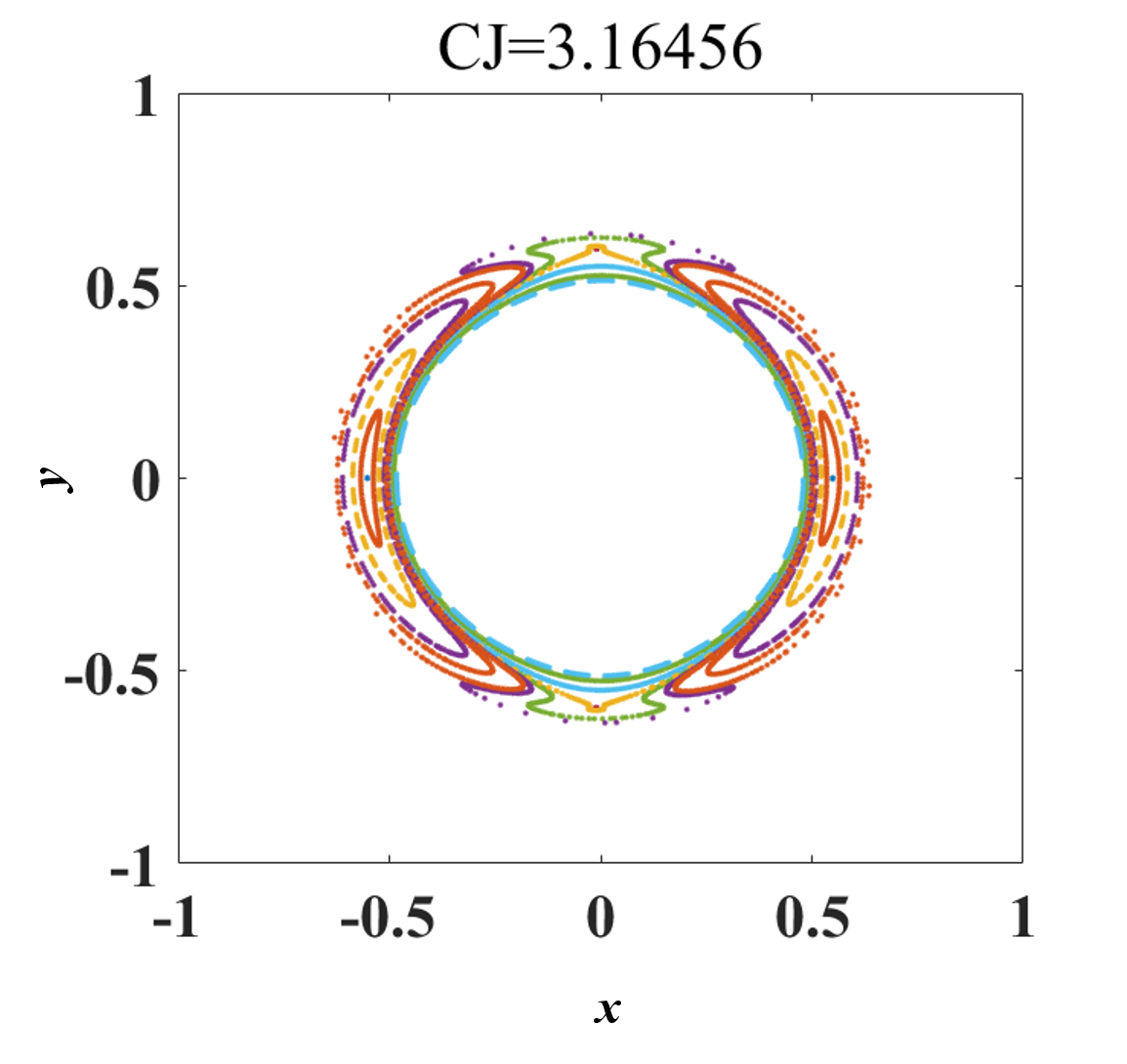}\\
   \includegraphics[width=2.5in]{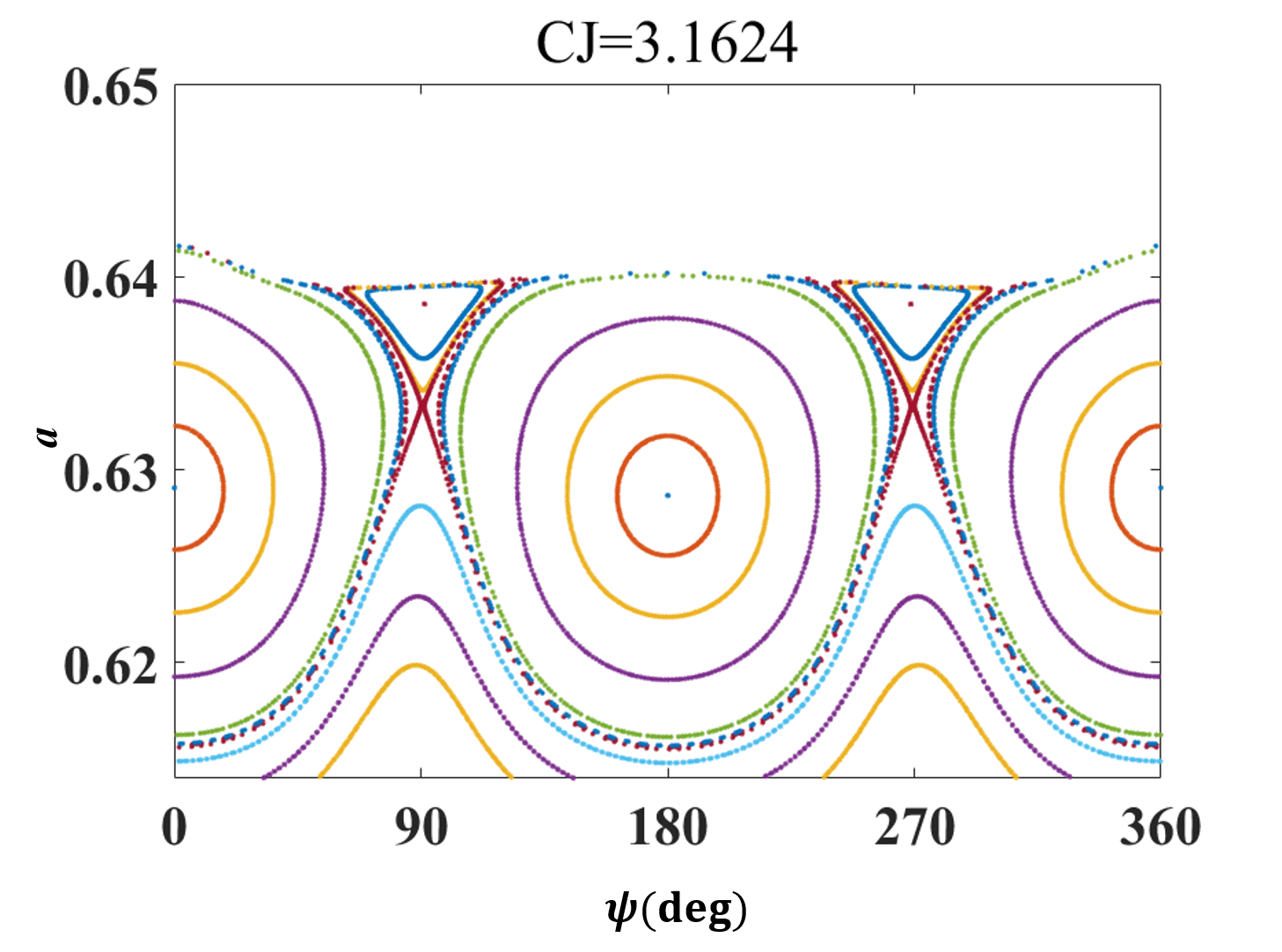} &
   \includegraphics[width=2in]{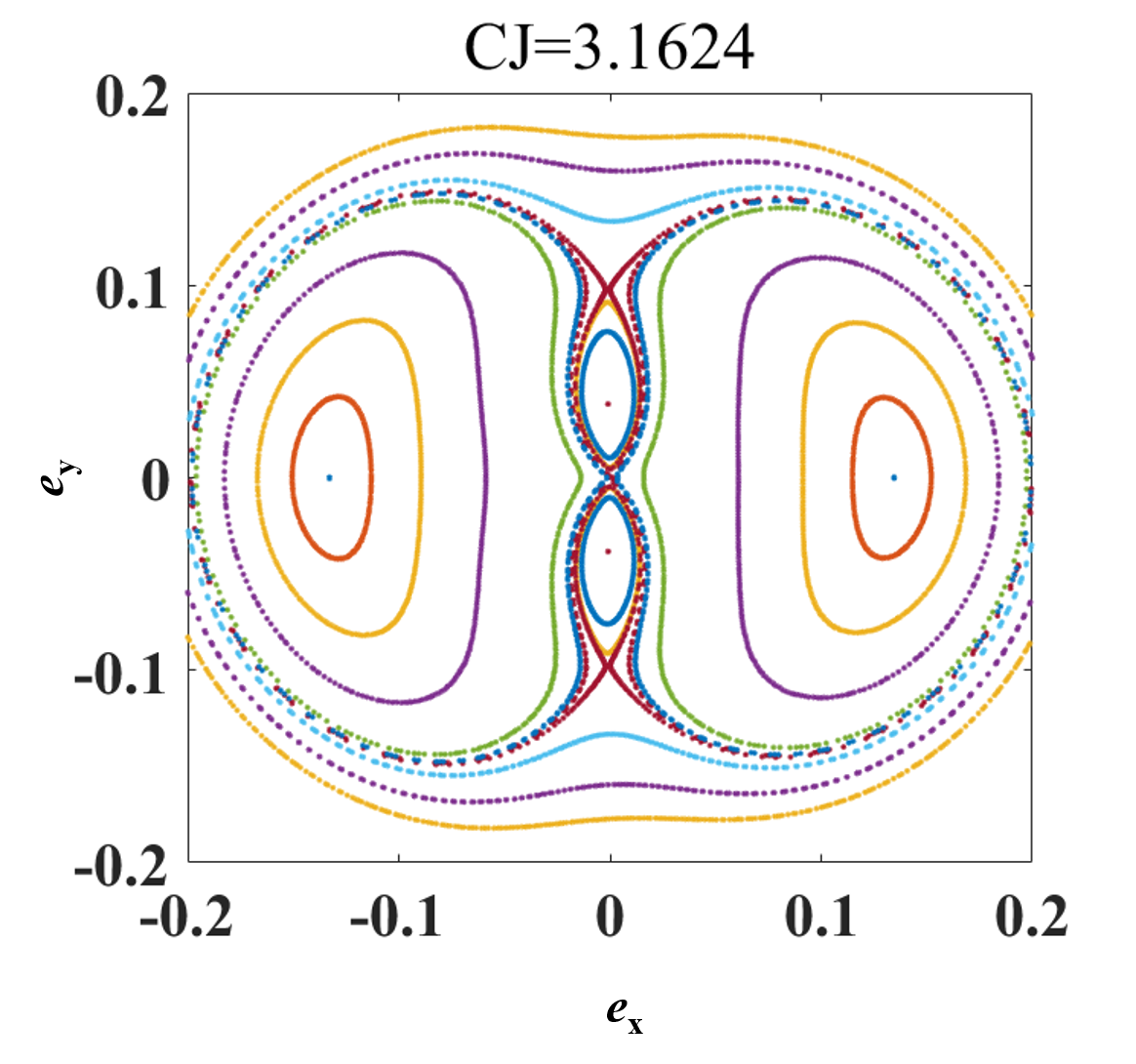} &
   \includegraphics[width=2in]{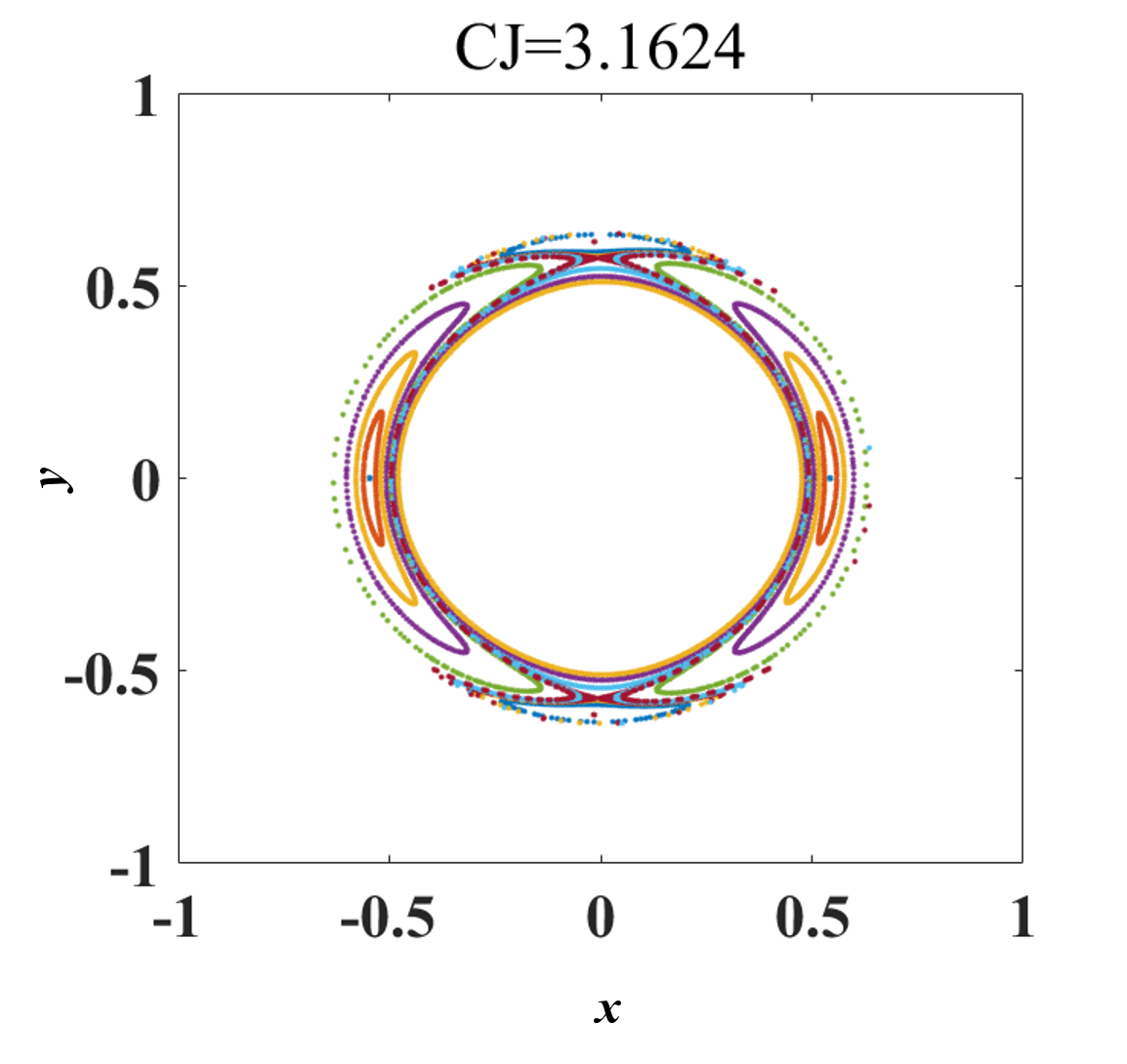}\\
   \includegraphics[width=2.5in]{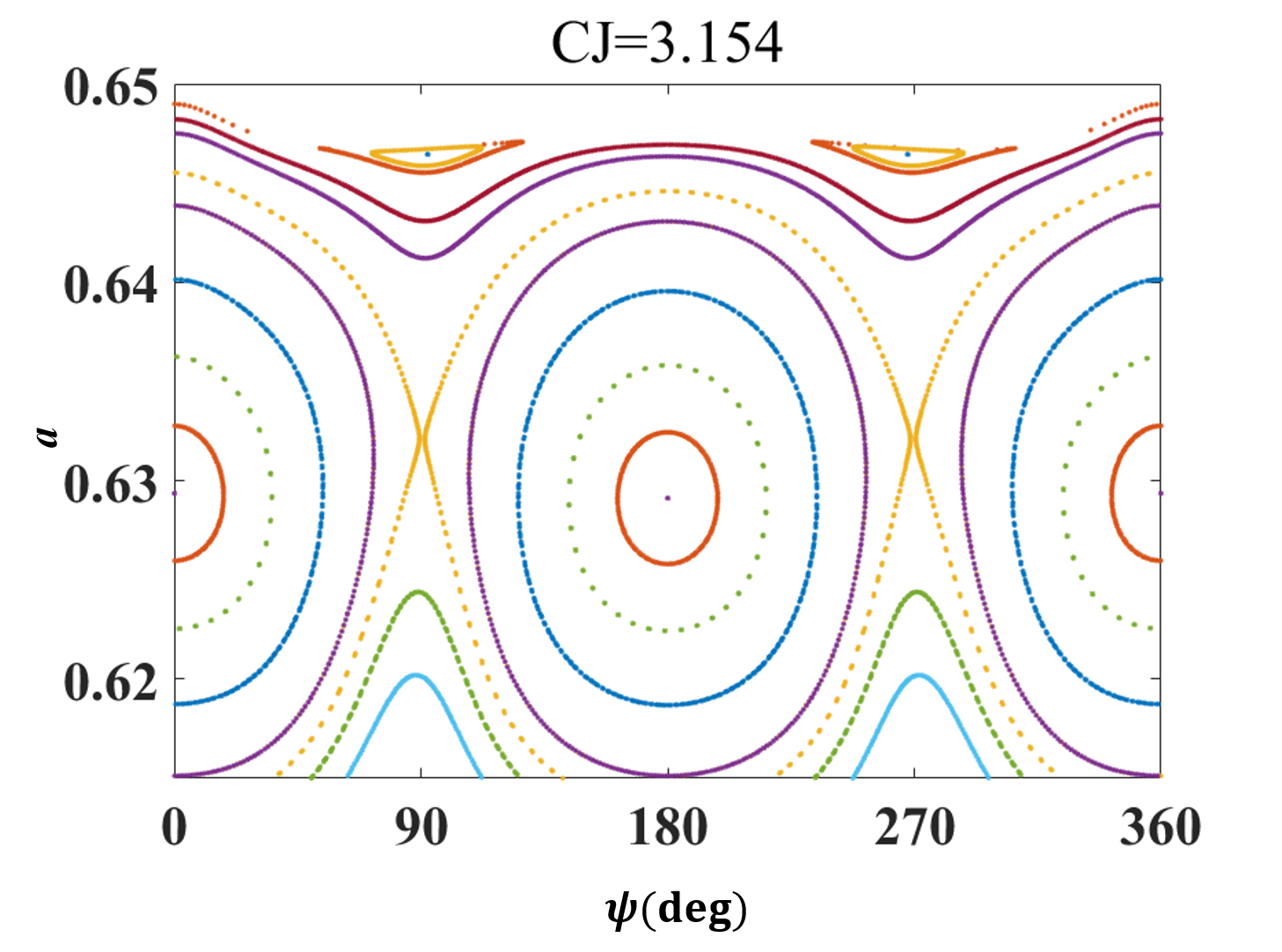} &
   \includegraphics[width=2in]{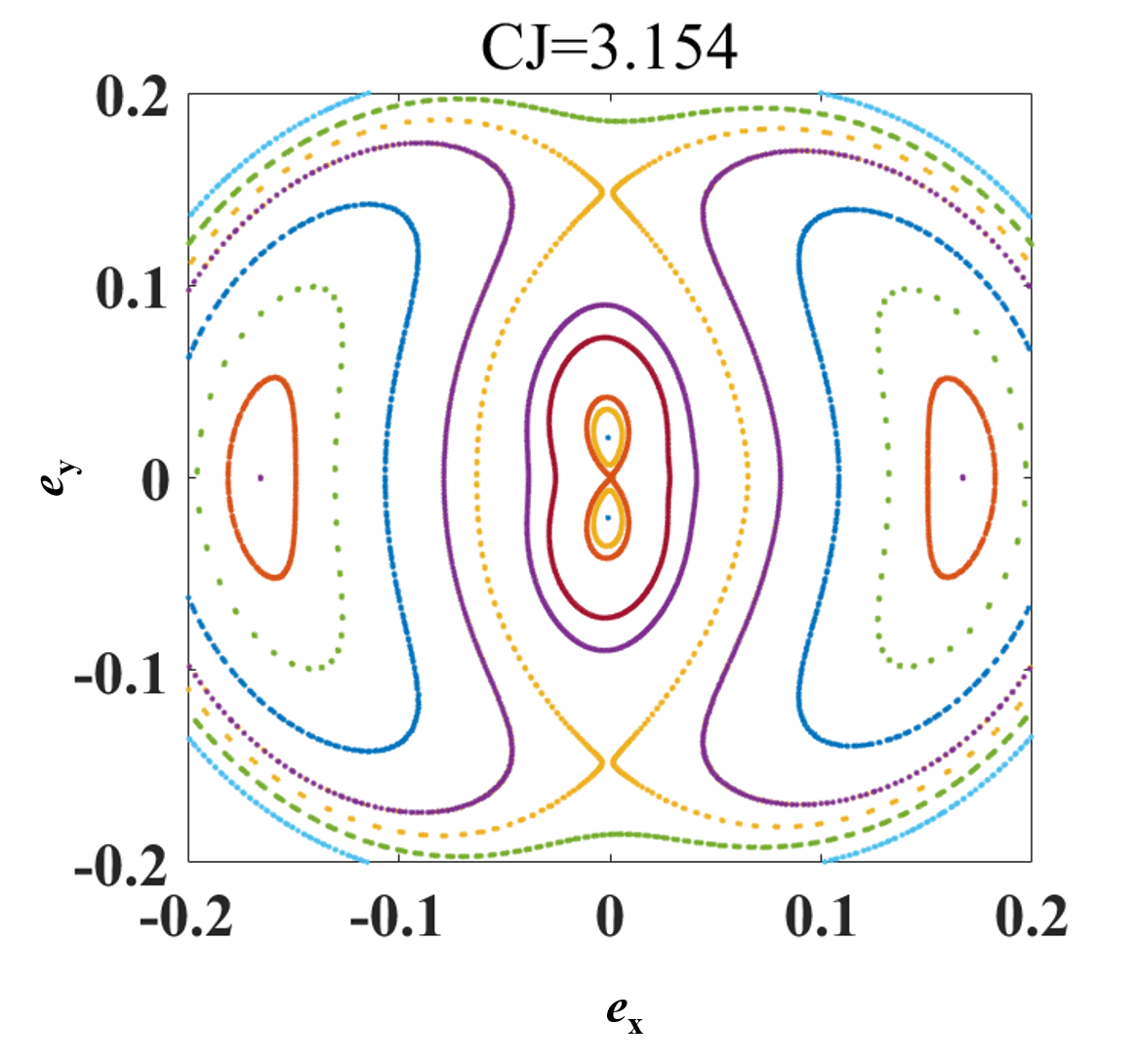} &
   \includegraphics[width=2in]{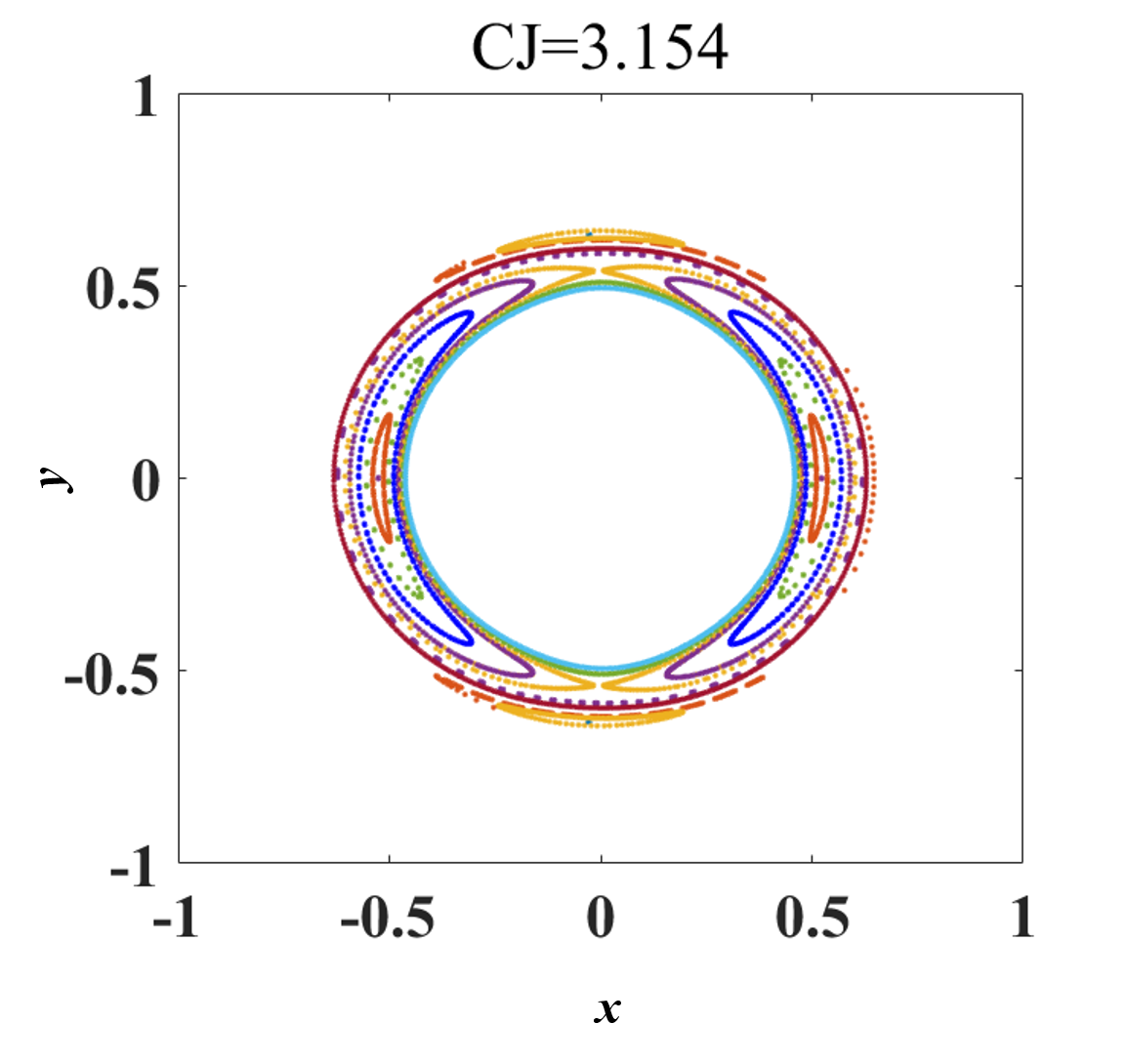}\\
   \end{tabular}   
 \caption{Poincar\'e sections near Jupiter's 2/1 interior resonance for three values of the Jacobi constant, decreasing from top to bottom, as indicated by the legend at the top of each panel. 
 The left column of panels displays the Poincar\'e sections in the $(\psi,a)$ plane, the middle column in the $(e_x,e_y)\equiv(e\cos\psi,e\sin\psi)$ plane. In the right panel we plot the Poincar\'e sections in configuration space, $(x,y)$, in the rotating frame; the locations of the Sun and Jupiter are fixed in this plane at $(x,y)=(-\mu,0)$ and $(x,y)=(1-\mu,0)$, respectively.}
 \label{f:f2}
\end{figure*}

The Poincar\'e sections can be plotted in several different variables; for our purposes, the most useful are plots of $(\psi,a)$, $(e\cos\psi,e\sin\psi)$ and $(x,y)$. Examples of these near Jupiter's 2/1 interior MMR are shown in Figure~\ref{f:f2}.  Consider the middle row in this figure, for a Jacobi constant value $C_J=3.1624$. Here in the left panel, we observe a prominent chain of two islands, each centered at $a\simeq0.629$ (slightly lower than the nominal exact resonant value, $a_{\rm res}=(1/2)^{2/3}\simeq0.630$) with $\psi=0$ and $\psi=180^\circ$. We also observe a less-prominent chain of two small islands, each centered at $a\simeq0.638$ (higher than the nominal exact resonant value), with $\psi\simeq90^\circ$ and $\psi\simeq270^\circ$. (The two separate chains of islands are discussed further in the next section.) In the middle and right panels of Figure~\ref{f:f2}, the same trajectories are projected in the $(e\cos\psi,e\sin\psi)$ plane and in the $(x,y)$ plane. 
The general properties of the behavior of test particles in the resonance are as follows. The test particle with initial conditions at the center of each chain of islands traces a periodic orbit in the four dimensional phase space which intersects the Poincar\'e section sequentially (and repeatedly) at each of the points at the island centers. The closed paths surrounding the centers of the islands are those test particles that trace quasi-periodic orbits librating about the exact periodic orbits, and the particle eccentricity and semi-major axis execute corresponding quasi-periodic variations. Most clearly visible in the middle panel is that each chain of islands is bounded by a separatrix beyond which $\psi$ circulates rather than librates.  
 
The 3/2 and 4/3 resonances exhibit qualitatively similar characteristics in their Poincar\'e sections, with the difference that the 3/2 MMR has two chains of three islands each while the 4/3 MMR has two chains of four islands each. The sizes of the stable libration islands are different for different values of the Jacobi constant. We measure the upper and lower boundaries in semi-major axis of each of the stable resonance islands in the $(\psi,a)$ Poincar\'e sections. As a function of particle eccentricity, we identify these measured resonance widths with the value of the eccentricity at the center of the corresponding islands in the $(e\cos\psi,e\sin\psi)$ plane. 

We generated many Poincar\'e sections for a range of values of the Jacobi constant to numerically compute the resonance widths over a almost the entire range of particle eccentricities, $0<e<1$ (emphasizing the small end of the range), in the neighborhood of the 2/1, 3/2 and 4/3 interior resonances of Jupiter. Then, with all the Poincar\'e sections in hand, we map the resonance zone boundaries in the $(a,e)$ plane. 
We describe the results in more detail in the following sections.

\section{Results}

\subsection{Two branches of the resonance at small eccentricity}

\begin{figure*}
 \centering
 \vglue-1.5in
 \includegraphics[angle=270,width=174mm]{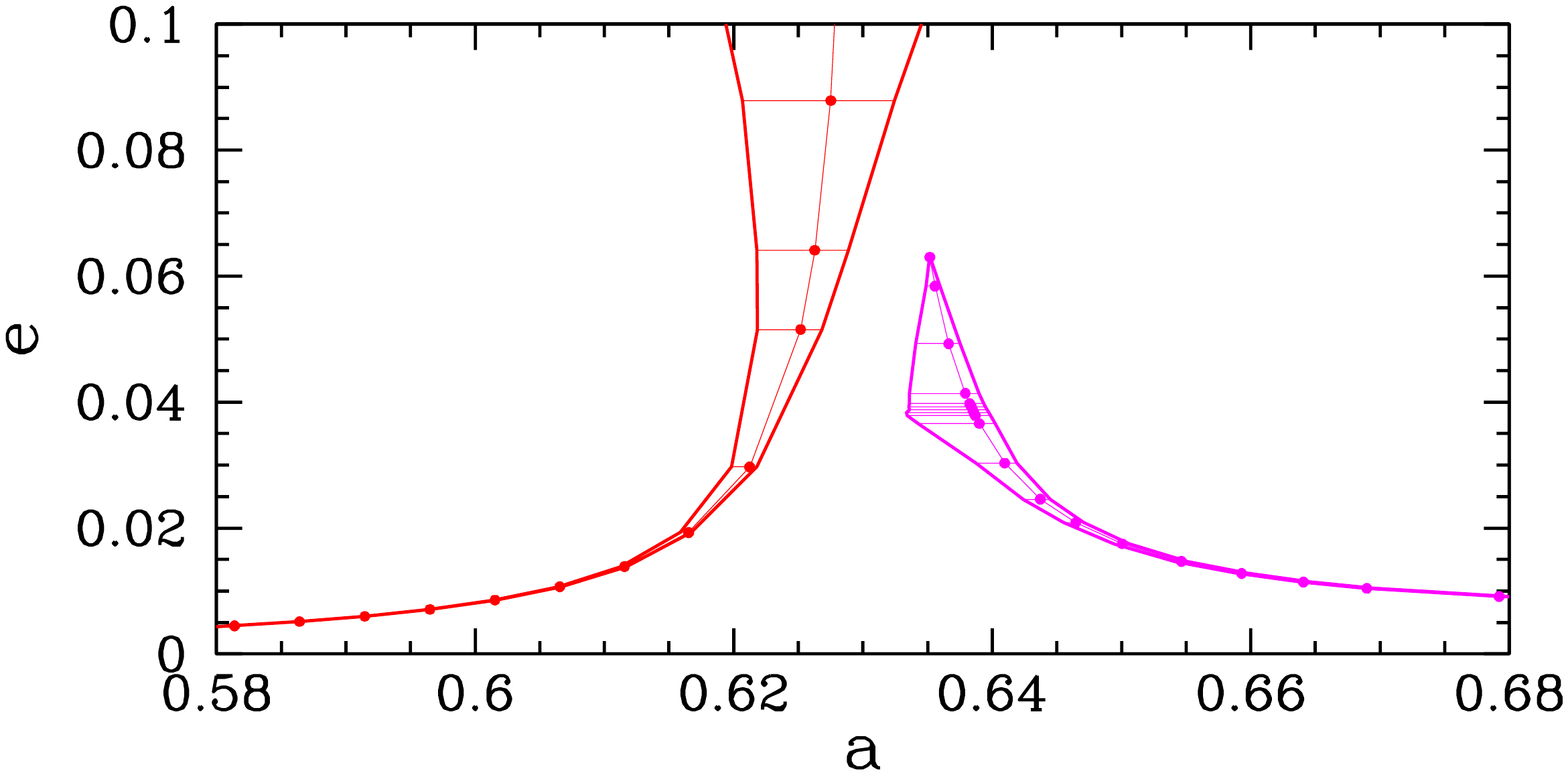}
 \caption{The libration center and width of Jupiter's 2/1 interior resonance for eccentricities in the range 0--0.1. The red curves (towards the left) indicate the pericentric libration zone centered at $\psi=180^\circ$ (equivalently, critical resonant angle $\phi$ centered at 0), and the magenta curves (towards the right) indicate the apocentric libration zone centered at $\psi=90^\circ$ (equivalently, critical resonant angle $\phi$ centered at 180$^\circ$). The points indicate the stable periodic orbit at the center of the corresponding resonant island, and the horizontal bars indicate the maximum range of librations of $a$. 
 }
 \label{f:f3}
\end{figure*}

The phase space structure near the 2/1 resonance is illustrated with three sets of Poincar\'e sections in Figure~\ref{f:f2}. Each row of panels displays the Poincar\'e section for a specific value of the Jacobi constant (increasing from the bottom to the top).
Focussing first on the middle row of panels in Figure~\ref{f:f2}, we observe two different pairs of stable islands, a pair that is prominent and large, and a second pair that is less prominent and small.  We call attention to a few important properties of these structures.
\begin{itemize}
\item The centers of the pair of large islands, at $(\psi,a)=(0,0.629)$ and $(\psi,a)=(180^\circ,0.629)$, describe a geometry of the periodic orbit in which the particle reaches perihelion alternately at conjunction ($\psi=0$) and at opposition ($\psi=180^\circ$) with Jupiter; the  conjunctions with Jupiter occur at alternate perihelion passages (i.e., at $\psi=0$). This can also be discerned in the $(x,y)$ plane shown in the panel on the right in which the large libration islands centered on the $x$-axis visualize these librations of the particle's perihelion in the rotating frame. (Figure 8.1 in \cite{Murray:1999SSD} illustrates this geometry.) In terms of the critical resonant angle (Eq.~\ref{e:phi-psi}), the corresponding libration zone is centered about $\phi=0$. We will call this the pericentric libration zone.
\item In the $(\psi,a)$ plane (left panel), we observe that the boundary of the large pair of stable islands is a separatrix that passes through a pair of unstable points at $\psi=90^\circ$ and $\psi=270^\circ$ near a value of $a\simeq0.635$. In the $(e_x,e_y)$ plane (middle panel), we see that this separatrix also passes through a third unstable point, at $e=0$, delineating the libration domain of the second pair of (smaller) stable islands at $\psi=90^\circ$ and $\psi=270^\circ$. 
\item The centers of the pair of small islands, at $(\psi,a)=(\pm90^\circ,0.639)$, as well as the nearby unstable pair of points, $(\psi,a)=(\pm90^\circ,0.635)$, describe a geometry in which the particle's conjunctions with Jupiter occur when the particle is near aphelion. (Figure 8.2 in \cite{Murray:1999SSD} illustrates this geometry; note however that those authors present this geometry as an unstable configuration, whereas we can see in the Poincar\'e sections that this configuration describes both stable and unstable periodic orbits, each of slightly different values of the particle's eccentricity and semi-major axis.)  In the $(x,y)$ section (right panel), we see that in this configuration, the particle's perihelion longitude librates alternately at $-90^\circ$ and $+90^\circ$ from Jupiter's longitude; this means that conjunctions with Jupiter occur near the particle's aphelion longitude.
In terms of the critical resonant angle (Eq.~\ref{e:phi-psi}), the stable periodic orbit at the center of the small islands has $\phi=180$, and the corresponding libration zone can be called the apocentric libration zone. 
\item The small pair of islands has a range in $\psi$ of only about $\pm30^\circ$ from each island's center, whereas the large pair of islands has a range in $\psi$ of $\pm90^\circ$. In terms of the critical resonant angle, the maximum libration amplitudes are $60^\circ$ and $180^\circ$ for the apocentric and pericentric librations, respectively.
\item For the pericentric libration zone, it is interesting to note that the island centered at $\psi=0$ has a range of semi-major axis slightly different than the one centered at $\psi=180$; this is due to the differences in the osculating elements at alternating perihelion passages of the test particle, one of which occurs at closer distance to Jupiter. 
\end{itemize}

Next, comparing the middle row of panels with the bottom and top rows in Figure~\ref{f:f2}, we observe qualitative and quantitative changes in the resonance structures for slightly different values of the Jacobi constant. 
In the bottom row, the pair of resonant islands at $\psi=0,180^\circ$ are larger than in the middle row, and centered at larger values of $e$; the pair of islands at $\psi=90^\circ,270^\circ$ are smaller than in the middle row, and centered at smaller values of $e$. In contrast with the middle panel, the separatrix bounding the large pair of islands in the bottom row does not pass through $e=0$, instead there is a second separatrix that passes though $e=0$ and it bounds the small pair of islands. In the top row, only the large pair of islands is visible, the small pair having nearly vanished at this value of the Jacobi constant. 

We recorded the locations of the stable resonance centers as well as the minimum and maximum values of the osculating semi-major axis $a$ for all the libration zone islands (as visible in the $(\psi,a)$ and the ($e\cos\psi,e\sin\psi)$ planes). 

In Figure~\ref{f:f3} we plot the resonance width of the 2/1 MMR in the $(a,e)$ plane. The plot shows the numerically determined resonance centers (and the corresponding resonance boundaries in semi-major axis) for one of each of the pair of islands representing the pericentric resonance zone and the apocentric resonance zone. For the pericentric zone, we plot the boundaries of the libration island centered about $\psi=180^\circ$. (The boundaries for the other libration island, centered at $\psi=0$, differ only very slightly from those of the island centered about $\psi=180^\circ$.) For the apocentric zone, we plot the boundaries centered at $\psi\approx90^\circ$. (The boundaries for the other island centered at $\psi=270^\circ$ are nearly indistinguishable from those of the island centered about $\psi=90^\circ$.) 
We observe that as the eccentricity approaches zero the pericentric resonance zone (shown with the red curves) shrinks and diverges towards the left of the figure (i.e., towards smaller values of semi-major axis) while the apocentric resonance zone (shown with the magenta curves) diverges towards the right of the figure (i.e., towards larger values of semi-major axis). The width of the pericentric zone shrinks rapidly as the eccentricity decreases. The width of the apocentric branch also decreases rapidly as the eccentricity decreases, but it is non-monotonic with eccentricity: it achieves a maximum width at an eccentricity of about $0.038$ and it vanishes at an eccentricity exceeding $\sim0.063$. At the very small eccentricities, the resonant librations far from the nominal resonant location are maintained by fast apsidal precession, with precession timescales of just a few times the orbital period.

In the pericentric libration zone (leftward of the nominal resonance location) the value of the particle's eccentricity at the center of the resonance increases rapidly as the semi-major axis of the resonance center approaches the nominal resonant value, $a_{\rm res}=(1/2)^{2/3}=0.630$. We can recognize that this behavior is related to the concept of "forced eccentricity" which diverges due to the small divisor in classical analytical perturbation theory for first order MMRs \citep[e.g.][]{Brouwer:1961,Murray:1999SSD}. In the context of planetary ring dynamics (e.g., a ring particle perturbed by a moon), this pericentric branch of the MMR is called a Lindblad resonance, a terminology apparently originating in galactic dynamics \citep{Murray:1999SSD}. 

The existence of two branches of the first order resonances is consistent with "the second fundamental model of resonance" developed by \cite{Henrard:1983} for isolated mean motion resonances at low eccentricities. 
A related perturbative analysis of the averaged planar elliptic restricted three body model by \cite{Henrard:1986}  showed that the effect of the perturber's eccentricity produces additional fine structure with the possibility of librations about the perturber's own pericenter.  Numerical analyses of Jupiter's interior 2/1 and 3/2 MMRs in the  planar elliptic restricted three body model confirmed the existence of such complexity and chaotic dynamics caused by the perturber's eccentricity \citep{Murray:1986}. In three-dimensional N-body numerical simulations, including the effects of Saturn and other planets, even more sources of chaotic dynamics have been identified within these resonances, such as secondary resonances between MMR librations and secular modes \citep{Morbidelli:1996,Ferraz-Mello:1999,Lecar:2001}. However, the associated chaotic diffusion timescales are very long, many orders of magnitude longer than the resonance libration periods, indicating that the underlying phase space structure based on the simplest PCRTB model remains relevant to understanding the dynamics.

\subsection{Bridges between neighboring first order MMRs}

\begin{figure*}
 \centering
 \vglue-1.5in\includegraphics[angle=270,width=174mm]{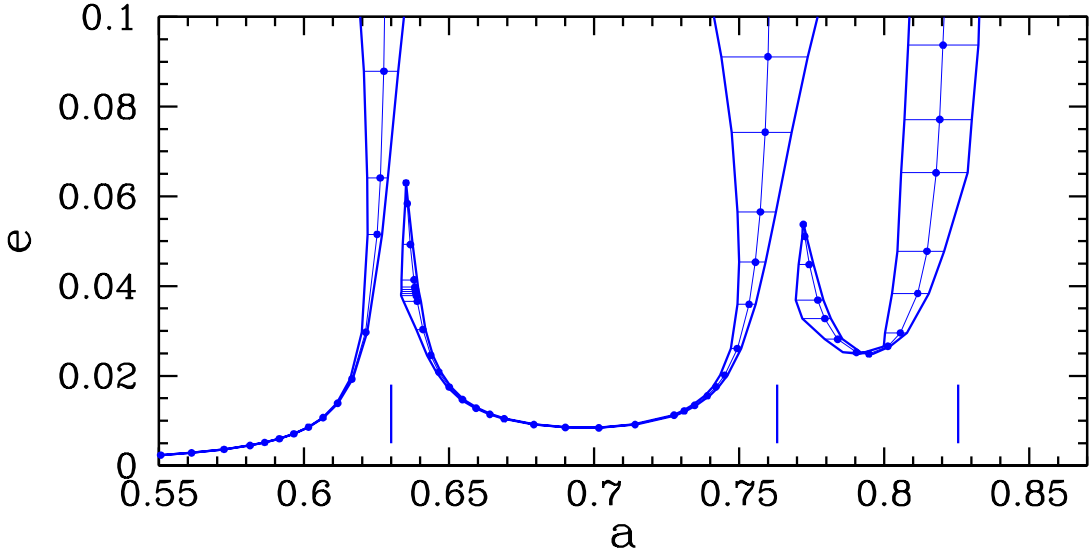}
 \caption{The libration centers and widths of Jupiter's 2/1, 3/2 and 4/3 interior resonances at low eccentricities, $0<e<0.1$. The nominal location of each resonance is indicated by the short vertical lines near the bottom of the figure. Each resonance has two branches, the pericentric zone (leftward of the nominal resonance location) and the apocentric zone (rightward of the nominal resonance location). The apocentric zone of the 2/1 resonance smoothly connects to the pericentric zone of the 3/2, and the apocentric zone of the 3/2 smoothly connects to the pericentric zone of the 4/3 resonance.}
 \label{f:f4}
\end{figure*}

\begin{figure*}
\centering
   \begin{tabular}{c c c}
   \includegraphics[width=2.5in]{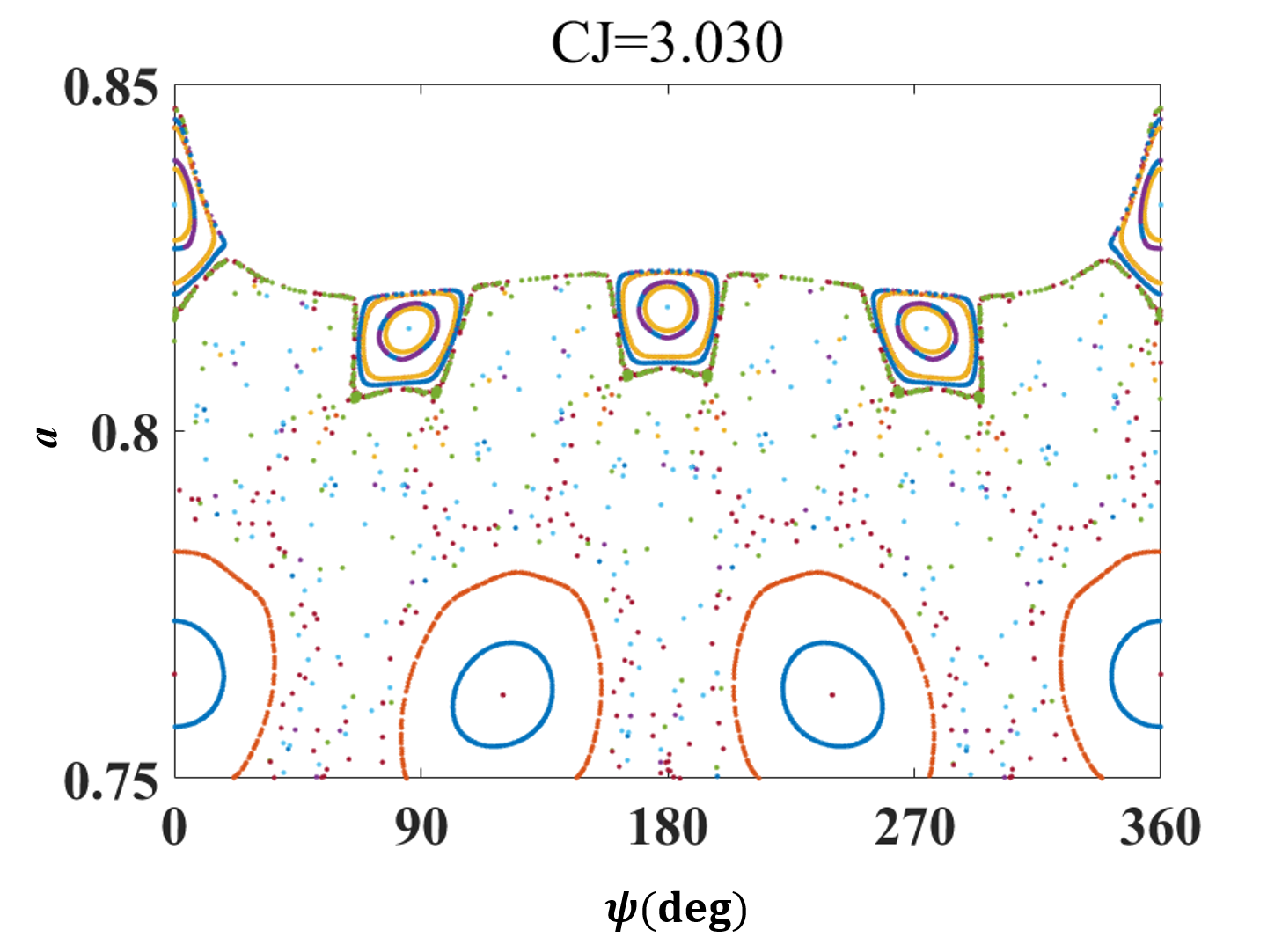} &
   \includegraphics[width=2in]{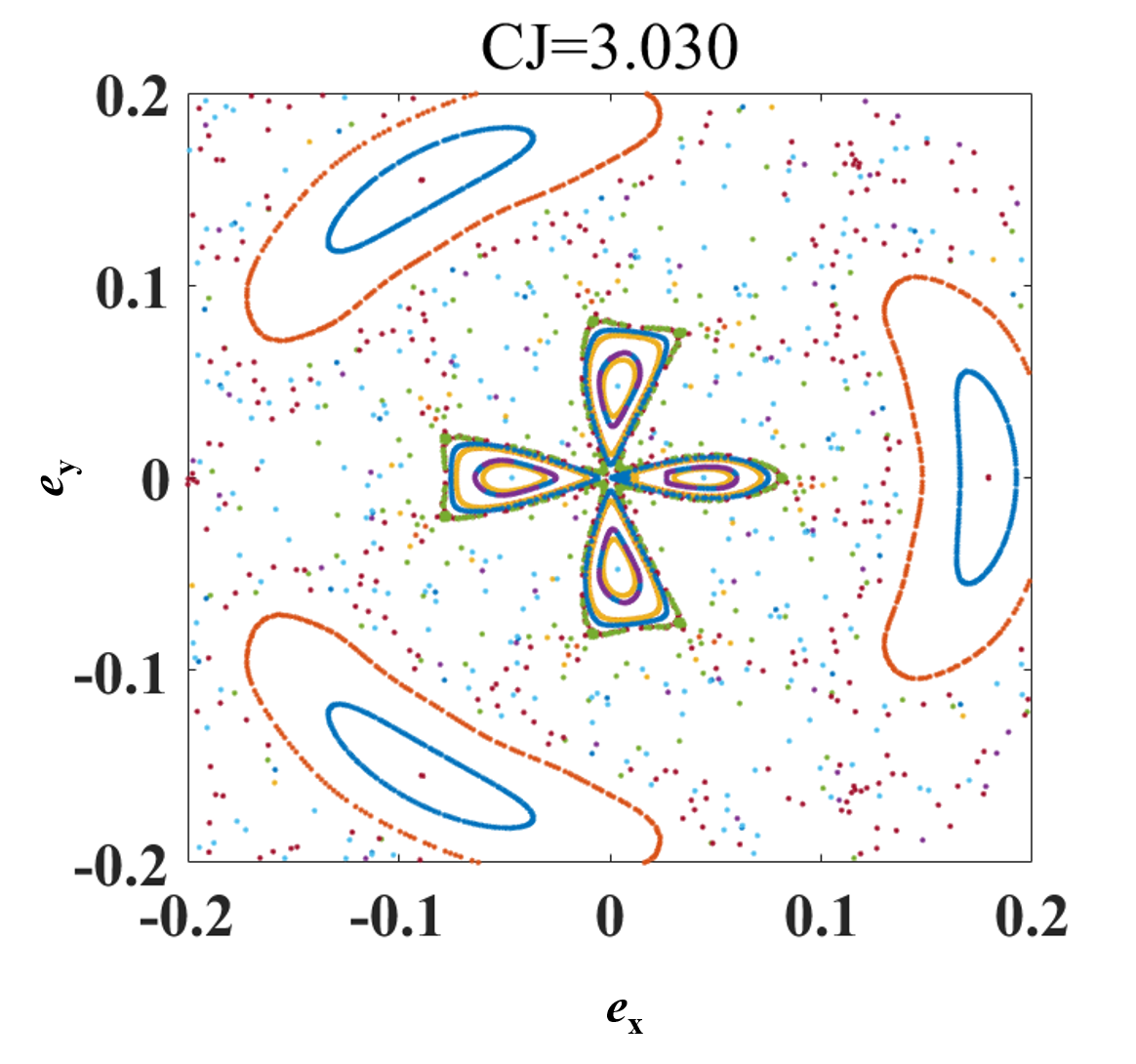} \\
      \includegraphics[width=2.5in]{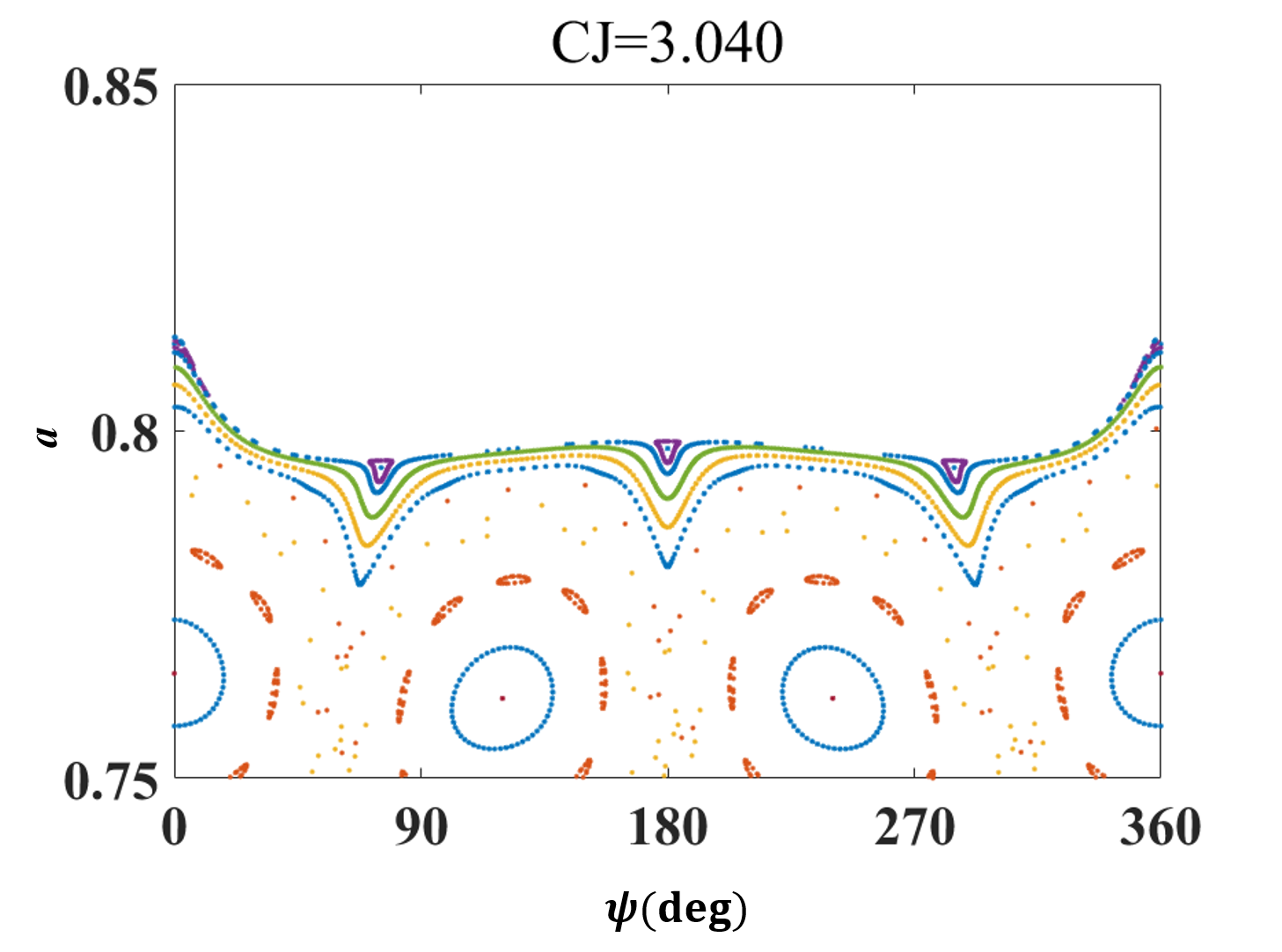} &
   \includegraphics[width=2in]{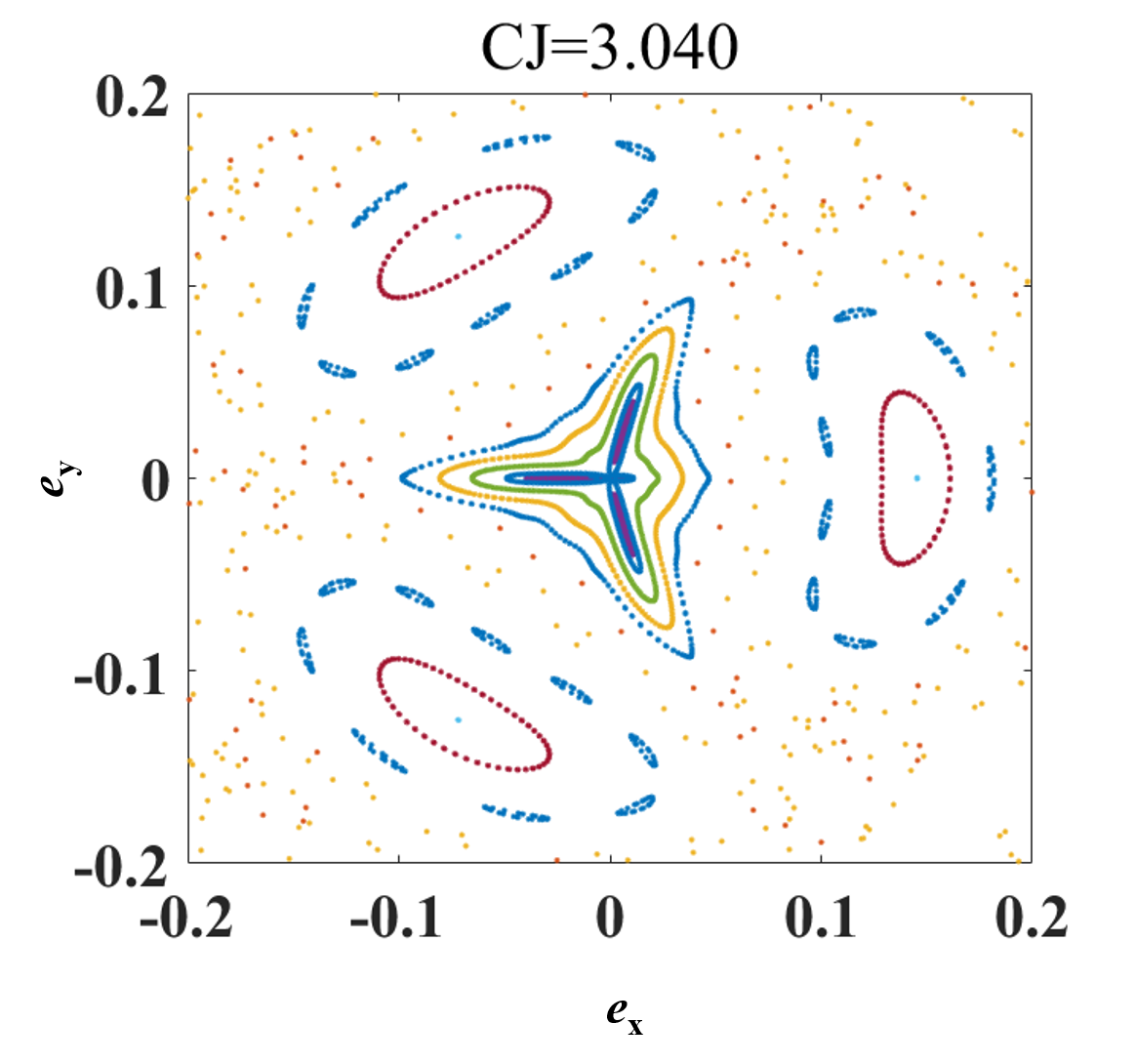} \\
      \includegraphics[width=2.5in]{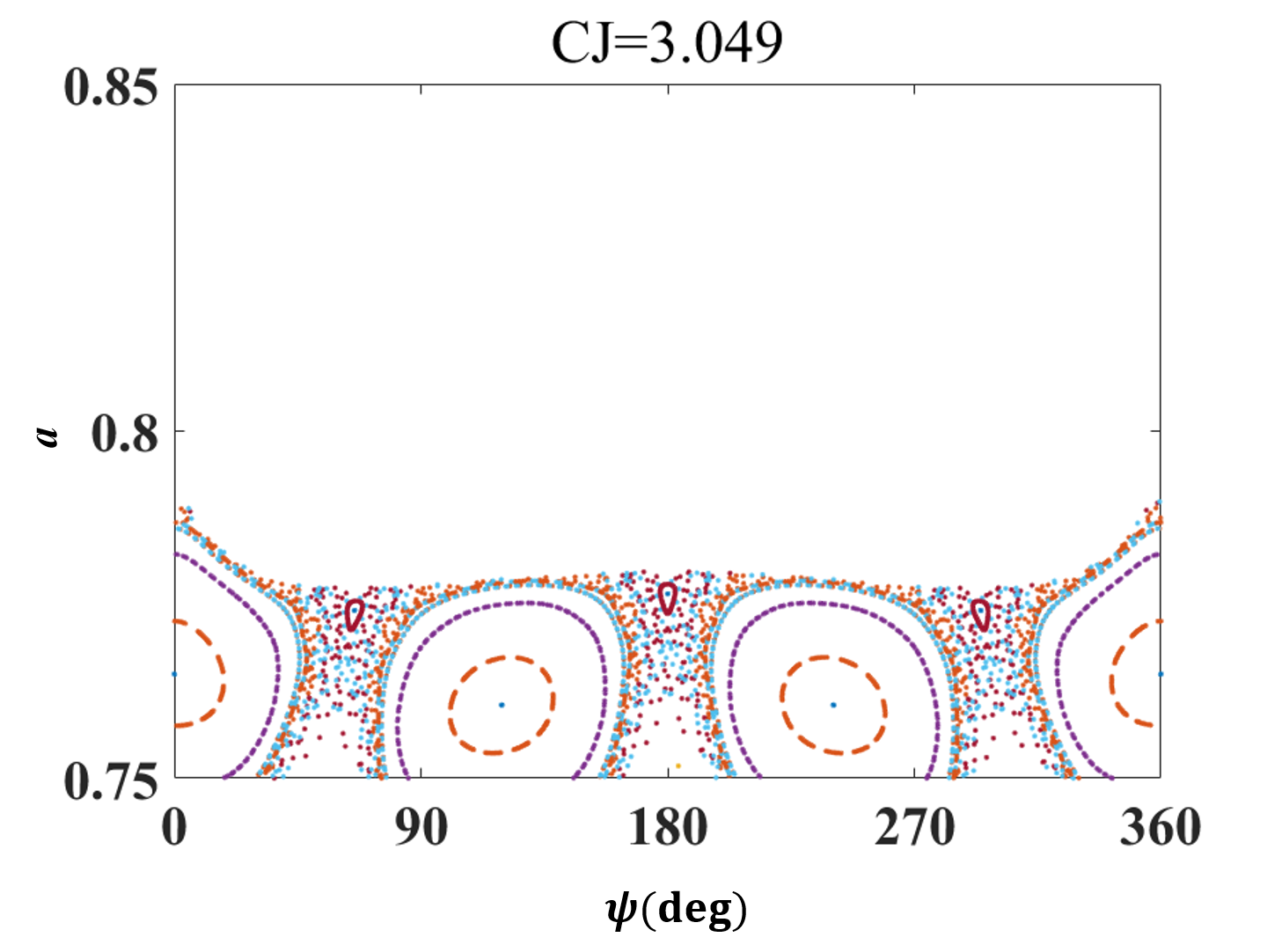} &
   \includegraphics[width=2in]{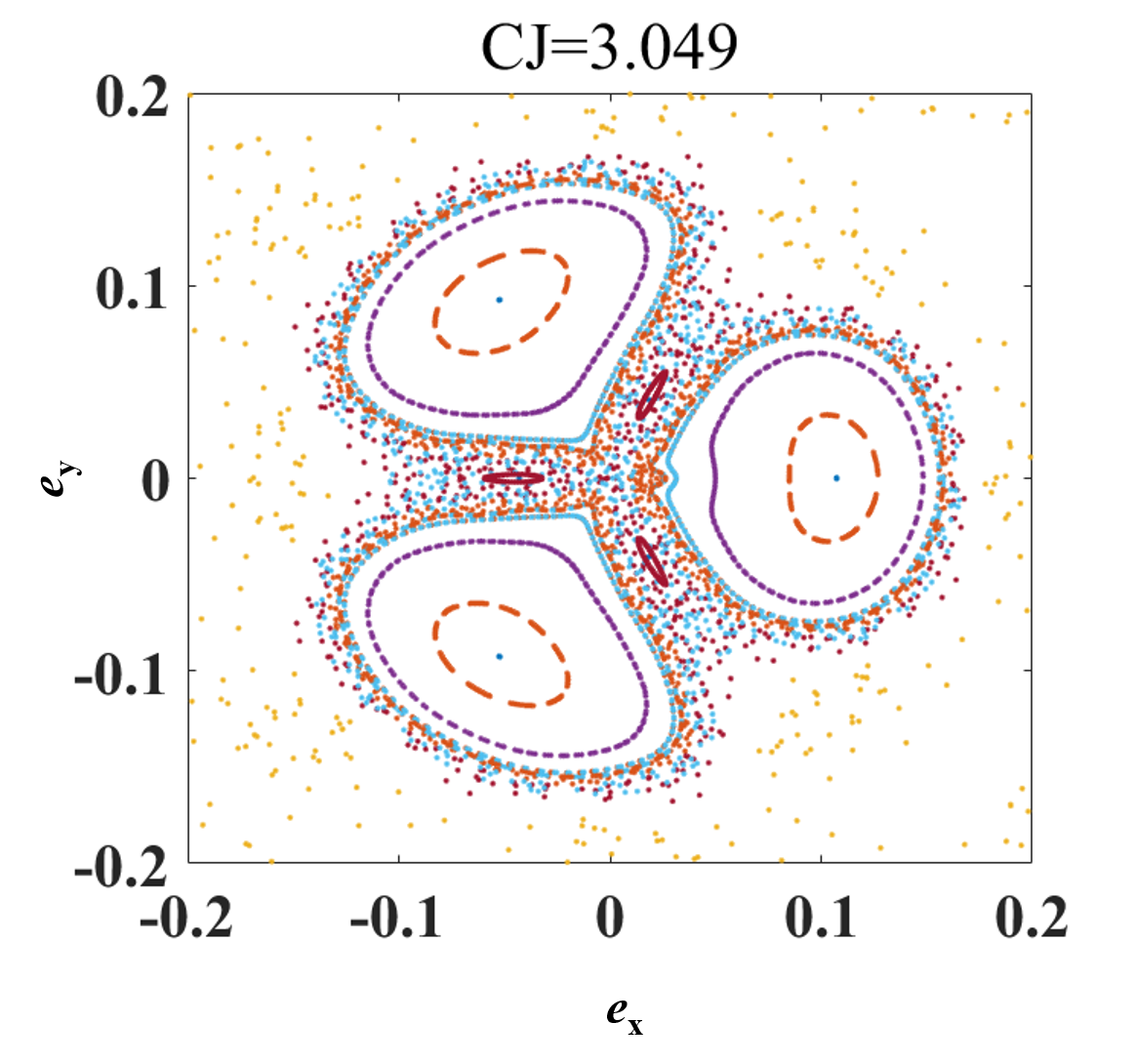} \\
   \end{tabular}   
 \caption{The phase space for low eccentricity orbits in proximity to the 4/3 and the 3/2 interior resonances of Jupiter. From the top row of panels to the bottom row of panels, the sequence of Jacobi constant values of 3.030, 3.040 and 3.049 traces the transition from the 4/3 pericentric resonance zone to the 3/2 apocentric resonance zone.  The left panels show the Poincar\'e sections in the $(a,\psi)$ plane; the projections of these same sections in the $(e\cos\psi,e\sin\psi)$ parameter plane are shown in the middle panel.
 }
 \label{f:f5}
\end{figure*}

We computed the resonance widths of the 3/2 and the 4/3 interior MMRs of Jupiter in a similar way as described above for the 2/1 MMR. These MMRs also have two branches of the resonance zone -- the pericentric zone and apocentric zone -- in the $(a,e)$ plane, analogous to those of the 2/1 MMR. In the Poincar\'e sections, the 3/2 MMR exhibits a chain of three libration islands centered at approximately $\psi=0,120,240$ for the pericentric branch, and another chain of three islands centered approximately at $\psi=60^\circ,180^\circ,300^\circ$ for the apocentric branch. The 4/3 MMR exhibits a chain of four libration islands centered approximately at $\psi=0,90^\circ,180^\circ,270^\circ$ for the pericentric branch, and at approximately $\psi=45^\circ,135^\circ,225^\circ,315^\circ$ for the apocentric branch. In Figure~\ref{f:f4} we show the boundaries in the $(a,e)$ plane of one representative libration island for each case. For the 2/1 MMR, the representative pericentric island is centered at $\psi=180$ and the representative apocentric island is centered at $\psi=90^\circ$. For the 3/2, the pericentric island shown is centered near $\psi=120^\circ$ and the apocentric island is centered near $\psi=60^\circ$. For the 4/3, the pericentric island shown is centered near $\psi=90^\circ$; the apocentric branch of the 4/3 is not shown because it has mostly dissolved into a chaotic zone for this choice of perturber mass.  

In examining  Figure~\ref{f:f4}, we observe a surprising feature: the apocentric branch of the 2/1 MMR smoothly transitions into the pericentric branch of the 3/2 MMR at very small eccentricities, and the apocentric branch of the 3/2 interior MMR smoothly transitions into the pericentric branch of the 4/3 interior MMR. In the $(a,e)$ plane these transitions present as low eccentricity "bridges" between neighboring first order MMRs.  

The mechanism of these transitions is illustrated in the sequence of Poincar\'e sections shown in Figure~\ref{f:f5}.  In this figure, the top row of Poincar\'e sections is for a Jacobi constant value $C_J=3.030$. Focussing on small eccentricities near the origin in the $(e\cos\psi,e\sin\psi)$ plane, we see a chain of four islands with approximately 4-fold symmetry, with centers at $e\approx0.03$; these are the resonant islands of the 4/3 MMR's pericentric resonance branch. 
This four-island chain is also visible near the upper boundary in the $(\psi,a)$ plane where we can observe that their centers are close to but slightly displaced from the nominal resonant value, $a_{\rm res}=(3/4)^{2\over3}=0.825$.  
In the second row in Figure~\ref{f:f5}, we show Poincar\'e sections for a slightly larger Jacobi constant value, $C_J=3.040$. Here we observe that the four-island chain near the origin of the $(e\cos\psi,e\sin\psi)$ plane has shrunk in size and its four-fold symmetry is significantly deteriorated: the island centered at $\psi=0$ is much smaller in size than the other three (just barely a "nub"), while the other three islands are larger, of similar size to each other, and they are closer to a three-fold symmetry. 
In the third row in Figure~\ref{f:f5}, we show Poincar\'e sections for a slightly even larger Jacobi constant value, $C_J=3.049$. Here we observe that the transformation of the four-island chain to a three-island chain is nearly complete: the island centered at $\psi=0$ has nearly vanished, and the other three islands (smaller in size than previously) are of similar size to each other, and they are visibly close to a three-fold symmetry. This three-island chain can be recognized as the apocentric branch of the 3/2 MMR whose centers are located at $\psi=60^\circ,180^\circ,300^\circ$ and at a slightly larger value of the semi-major axis than the nominal resonant value $a_{\rm res}= (2/3)^{2\over3}=0.763$.

We see that in this transition one of the four pericentric libration islands of the 4/3 MMR, the one centered at $\psi=0$, vanishes at some small eccentricity, while the other three pericentric libration islands centered near $\psi=90^\circ,180^\circ,270^\circ$ gradually evolve into the three apocentric libration islands of the 3/2 MMR centered near $\psi=60^\circ,180^\circ,240^\circ$, respectively.  This means that when we represent the resonance widths in the $(a,e)$ plane, only three of the four pericentric libration zones of the 4/3 MMR continuously evolve into the three apocentric libration zones of the 3/2 MMR at low eccentricities. Also to be noted, the 4/3 MMR's two islands centered near $\psi=90^\circ,270^\circ$ (which evolve into the 3/2 MMR's apocentric islands centered near $\psi=60^\circ,240^\circ$, respectively) have the same trace in the $(a,e)$ plane, but the other two islands, centered at $\psi=0,180^\circ$ trace differently in the $(a,e)$ plane.

Similarly, in the low eccentricity regime, only two of the three pericentric libration islands of the 3/2 MMR (the two centered near $\psi=120^\circ,240^\circ$) evolve into the two apocentric libration islands of the 2/1 MMR (those centered near $\psi=90^\circ,270^\circ$, respectively). The third pericentric island of the 3/2 MMR, the one centered at $\psi=0$, vanishes as the 3/2 pericentric zone evolves into the chain of two islands of the apocentric branch of the 2/1 MMR.

We mention a notable difference between the 2/1 MMR versus the 3/2 and 4/3 MMRs which is obvious in the Poincar\'e sections: the latter have visible and significant chaotic zones at the resonance boundaries, whereas no chaotic zones are visible for the 2/1 MMR at low eccentricities. The chaotic zones are anticipated from the resonance overlap criterion for conservative dynamical systems \citep{Chirikov:1979} applied to first-order $(p+1)/p$ MMRs at low eccentricity of the test particle \citep{Wisdom:1980} which predicts resonance overlap leading to chaos for $p\gtrsim 0.51\mu^{-{2\over7}}$, i.e., $p\gtrsim4$ for the value of $\mu$ for Jupiter-Sun. 

\subsection{Resonance widths over the full range of eccentricity}

\begin{figure*}
 \centering
  \vglue-1.5in\includegraphics[angle=270,width=7in]{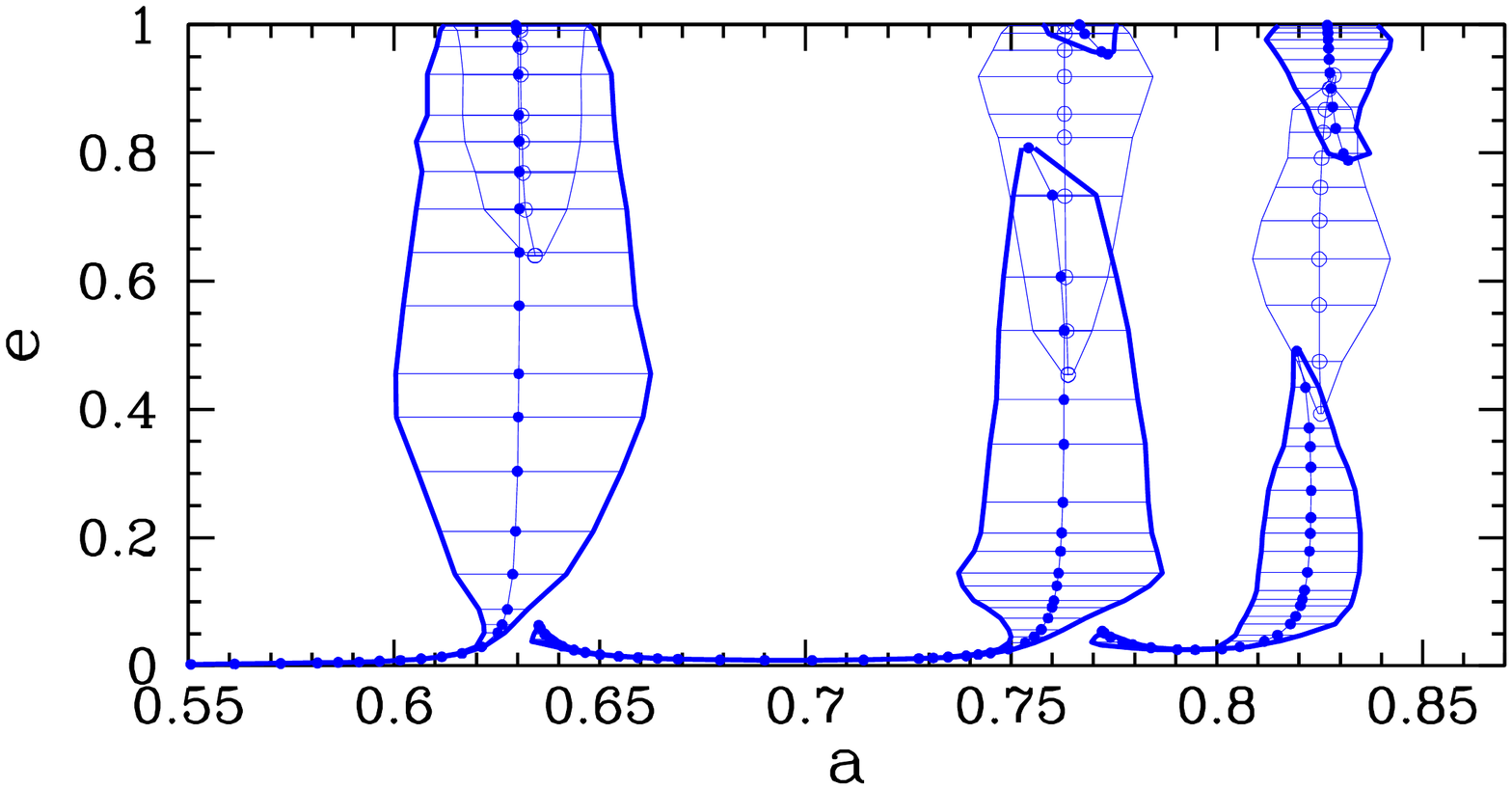}
 \caption{The libration centers and widths of Jupiter's 2/1, 3/2 and 4/3 interior resonances for the full range of eccentricities, $0<e<1$, for representative pericentric and apocentric libration islands, in each case. The center of the pericentric libration island is indicated by the filled circles, that of the apocentric libration island is indicated by the open circles.  The maximum libration range of $a$ is indicated by the horizontal bars and thick curves (pericentric zone) and thin curves (apocentric zone).
}
 \label{f:f6}
\end{figure*}

The primary focus of this work has been the phase space structure and widths of first order resonances at low eccentricities. For completeness, we undertook to measure the resonance widths of Jupiter's interior 2/1, 3/2 and 4/3 MMRs at higher eccentricities as well. Figure~\ref{f:f6} plots these widths in the $(a,e)$ plane for the full range, $0<e<1$. 

At higher eccentricities the center of the pericentric zone closely follows the nominal resonance location at $a_{\rm res}=((p+1)/p)^{-{2\over3}}$. Its width however is not monotonically increasing with eccentricity, as the pendulum model for MMRs would indicate. Rather, the pericentric zone width achieves a maximum and then decreases with increasing eccentricity. In the case of the 3/2 and 4/3 MMRs, the pericentric zone vanishes near eccentricity  0.8 and 0.5, respectively, and reappears again at eccentricity near 0.95 and 0.80, respectively. 

The apocentric libration zone also exhibits several discontinuities. As mentioned above, at low eccentricities, this zone exists only for eccentricities below a critical value, $e_{\rm crit}$.  But it reappears at eccentricities above the planet-crossing value, $e_c = a_{\rm res}^{-1} - 1$; moreover, for the 4/3 MMR, it vanishes and reappears multiple times at even higher eccentricities. 

These discontinuities at high eccentricities were explained by \cite{Wang:2017} with reference to the shape of the 3/2 resonant orbit in the rotating frame; a similar explanation applies to the case of the 4/3 MMR. (Resonance widths at high eccentricities for many exterior resonances have been investigated in \cite{Lan:2019}, and these also show multiple discontinuities.) The reason for these discontinuities is that at high eccentricities the trace of the resonant orbit in the rotating frame cuts the circle of radius $1-\mu$ at multiple points, giving rise to new pairs of stable and unstable equilibrium points in phase space. 

\begin{figure*}
 \centering
 \vglue-1.5in \includegraphics[angle=270,width=7in]{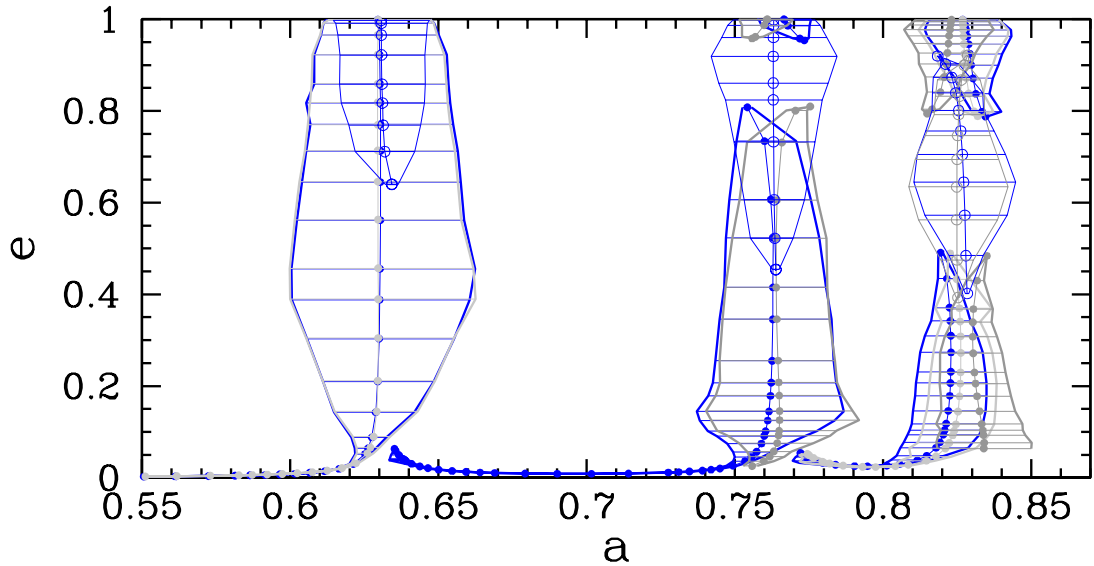}
 \caption{Similar to Fig.~\ref{f:f6}, but including additional libration islands for each MMR; these are over-plotted in gray.
}
 \label{f:f7}
\end{figure*}

Also worthy of mention are some differences amongst the phase space structures of the 2/1, 3/2 and 4/3 MMRs. As mentioned previously, in the case of the 2/1 MMR, the Poincar\'e sections have an approximate two-fold symmetry of the two libration islands of the pericentric branch (respectively, apocentric branch) and are only slightly different from each other in their presentation in the $(a,e)$ plane.  However, the three pericentric (respectively, apocentric) libration islands of the 3/2 MMR depart more noticeably from three-fold symmetry. Their libration centers and boundaries in the $(a,e)$ plane are also more noticeably different from each other. Even greater deviations (from four-fold symmetry) are found amongst the four libration islands of the pericentric branch of the 4/3 MMR, leading to correspondingly greater differences in the $(a,e)$ resonance boundaries of the different libration centers.  These are plotted in Figure~\ref{f:f7}.

\section{Summary and Discussion}

We undertook a non-perturbative computation of the resonance widths at low eccentricities of first order MMRs, specifically Jupiter's interior 2/1, 3/2, and 4/3 MMRs in the planar circular restricted three body model. Our approach is numerical and uses Poincar\'e sections to measure the resonance libration zone boundaries in the numerically computed $(a,\psi)$ section where $a$ is the osculating semi-major axis and $\psi$ is the longitude of conjunction relative to the longitude of perihelion. By computing Poincar\'e sections for a range of Jacobi constant values, we investigated the resonance structures for a range of eccentricities. For small eccentricities, $e<0.1$, we obtained the following results for the resonance widths traced in the $(a,e)$ plane.

\begin{enumerate}

\item First order interior MMRs present two distinct libration zones, the "pericentric" and "apocentric" librations.  The center of the pericentric libration zone occurs at $a<a_{\rm res}$ whereas that of the apocentric libration zone occurs at $a>a_{\rm res}$, where $a_{\rm res}=(p/(p+1))^{{2\over3}}$ is the nominal location of the $(p+1)/p$ MMR.  As the eccentricity decreases, the resonance centers diverge away from $a_{\rm res}$ and their widths shrink (Fig.~\ref{f:f3} and Fig~\ref{f:f4}). 

\item The apocentric libration zone is discontinuous: it exists only for eccentricities below a critical value, $e_{\rm crit}$. From our numerical analysis for Jupiter's MMRs, we find $e_{\rm crit} = 0.063$ and $0.055$ for the 2/1 and the 3/2 MMRs, respectively. By referring to analytical results \citep{Henrard:1983,Dermott:1988}, we infer that $e_{\rm crit}$ scales with perturber mass, $\mu$, and MMR integer, $p$, as $\sim(\mu/p)^{1\over3}$.

\item Neighboring first order MMRs are connected with "low eccentricity bridges" (Fig.~\ref{f:f4}): the apocentric libration zone of the $(p+1)/p$ interior MMR smoothly transitions into the pericentric libration zone of the $(p+2)/(p+1)$ interior MMR. Only the 2/1 MMR has a pericentric libration zone apparently not connected to any other MMR; for $e$ approaching zero, its location diverges to small values of $a$. (However, we did not investigate very low values of $a$ and we cannot rule out bridges with closer-in resonance structures, including the 1/1 MMR and the Lagrange points.)

\item In terms of the critical resonance angle, $\phi$, the maximum libration amplitude of the pericentric librations is up to $180^\circ$, but that of the apocentric librations does not exceed $60^\circ$. Only Jupiter's 2/1 MMR reaches the maximum libration amplitudes (see Fig.~\ref{f:f2}); the 3/2 and the 4/3 MMRs under-achieve these maximum ranges, in part because their resonance separatrices broaden into chaotic zones even at low eccentricities, $e<0.1$ (Fig.~\ref{f:f5}). 

\end{enumerate}

We mention a few other notable features that emerge from our non-perturbative analysis that are different from the analytical perturbative treatments and numerical averaging approaches in the literature. These features reveal the limitations of the analytical and numerical averaging approaches.
\begin{enumerate}
\item The resonance separatrix does not vanish at low eccentricities. It is very much evident in the Poincar\'e sections where we see that the small eccentricity apocentric libration zone is delineated by its own separatrix in the $(e\cos\psi,e\sin\psi)$ plane (Figure~\ref{f:f2}). This feature has not been found in previous analyses which have been based on the use of the critical resonance angle, $\phi$. We think that our choice of $\psi$ for visualizing the  Poincar\'e sections is important in revealing the existence of the separatrix at low eccentricities. 
\item At very low eccentricities, the resonance center is visibly displaced from the nominal values of semi-major axis, $a_{\rm res}=(p/(p+1))^{2\over3}$. The chain of libration islands also deviate significantly from $(p+1)$-fold symmetry. These are both owed to the gradual evolution of the apocentric zone of the $(p+1)/p$ MMR into the pericentric zone of the neighboring $(p+2)/(p+1)$ MMR in the low eccentricity bridges between first order MMRs. The deviations from $(p+1)$-fold symmetry are greater for larger values of $p$. 
\item The low eccentricity bridges between adjacent first order MMRs are understandably not revealed in previous analyses which treat every MMR in isolation to build a single-resonance theory, either with analytical perturbation theory or with numerical averaging. 
\end{enumerate}

It is noteworthy that the apocentric libration zone and the low eccentricity bridges dissolve into chaotic zones in close proximity to the perturber. We found numerically that this occurs for the 4/3 MMR in the Jupiter-Sun case, consistent with the analytical criterion for first order resonance overlap \citep{Wisdom:1980}. We can anticipate that for smaller $\mu$, the dissolution of the low eccentricity bridges would occur at even closer proximity to the perturber. For example, for $\mu=3\times10^{-5}$ (a mass ratio common in exo-planetary systems and amongst the regular satellite systems of the giant planets, and similar to that of Neptune/Sun), the low eccentricity bridges could exist for many more first order $(p+1)/p$ MMRs, up to $p\approx10$. 

Our non-perturbative investigation has revealed several aspects of the fine structure of first order MMRs at low eccentricities not found in previous studies. 
Our novel finding of low eccentricity bridges between first order resonances has implications for radial transport in planetary systems, provided these bridges are strong enough to persist under the additional perturbations present in real systems. For example, under external dissipative effects (solar radiation forces, solar mass loss, gas drag, tidal dissipation, etc.), adiabatic resonant migration along the low eccentricity bridges may effect transport across large radial distances in planetary systems. This has implications for our current understanding of the radial mixing and provenance of small body populations as well as resonant capture and migration of planets and satellites, all in the low eccentricity regime. We leave these interesting applications to a future investigation.

\section*{Data Availability}
The data (or codes to generate the data) underlying the figures in this article are available at \url{https://github.com/renumalhotra/2020-first-order-mmrs}.

\section*{Acknowledgements}
We thank an anonymous reviewer for comments that improved this paper. RM is grateful for research support from NSF (grant AST-1824869), NASA (grant 80NSSC18K0397) and the Louise Foucar Marshall Foundation. NZ acknowledges funding from the National Natural Science Foundation of China for Distinguished Young Scholars (No.~11525208).




\bibliographystyle{mnras}

\begin{thebibliography}{}
\makeatletter
\relax
\def\mn@urlcharsother{\let\do\@makeother \do\$\do\&\do\#\do\^\do\_\do\%\do\~}
\def\mn@doi{\begingroup\mn@urlcharsother \@ifnextchar [ {\mn@doi@}
  {\mn@doi@[]}}
\def\mn@doi@[#1]#2{\def\@tempa{#1}\ifx\@tempa\@empty \href
  {http://dx.doi.org/#2} {doi:#2}\else \href {http://dx.doi.org/#2} {#1}\fi
  \endgroup}
\def\mn@eprint#1#2{\mn@eprint@#1:#2::\@nil}
\def\mn@eprint@arXiv#1{\href {http://arxiv.org/abs/#1} {{\tt arXiv:#1}}}
\def\mn@eprint@dblp#1{\href {http://dblp.uni-trier.de/rec/bibtex/#1.xml}
  {dblp:#1}}
\def\mn@eprint@#1:#2:#3:#4\@nil{\def\@tempa {#1}\def\@tempb {#2}\def\@tempc
  {#3}\ifx \@tempc \@empty \let \@tempc \@tempb \let \@tempb \@tempa \fi \ifx
  \@tempb \@empty \def\@tempb {arXiv}\fi \@ifundefined
  {mn@eprint@\@tempb}{\@tempb:\@tempc}{\expandafter \expandafter \csname
  mn@eprint@\@tempb\endcsname \expandafter{\@tempc}}}

\bibitem[\protect\citeauthoryear{{Brouwer} \& {Clemence}}{{Brouwer} \&
  {Clemence}}{1961}]{Brouwer:1961}
{Brouwer} D.,  {Clemence} G.~M.,  1961, {Methods of celestial mechanics}

\bibitem[\protect\citeauthoryear{{Chirikov}}{{Chirikov}}{1979}]{Chirikov:1979}
{Chirikov} B.~V.,  1979, \mn@doi [Physics Reports]
  {10.1016/0370-1573(79)90023-1}, \href
  {http://adsabs.harvard.edu/abs/1979PhR....52..263C} {52, 263}

\bibitem[\protect\citeauthoryear{{Deck}, {Payne}  \& {Holman}}{{Deck}
  et~al.}{2013}]{Deck:2013}
{Deck} K.~M.,  {Payne} M.,   {Holman} M.~J.,  2013, \mn@doi [\apj]
  {10.1088/0004-637X/774/2/129}, \href
  {https://ui.adsabs.harvard.edu/abs/2013ApJ...774..129D} {774, 129}

\bibitem[\protect\citeauthoryear{{Dermott}, {Malhotra}  \& {Murray}}{{Dermott}
  et~al.}{1988}]{Dermott:1988}
{Dermott} S.~F.,  {Malhotra} R.,   {Murray} C.~D.,  1988, \mn@doi [\icarus]
  {10.1016/0019-1035(88)90074-7}, \href
  {http://adsabs.harvard.edu/abs/1988Icar...76..295D} {76, 295}

\bibitem[\protect\citeauthoryear{Fehlberg}{Fehlberg}{1968}]{Fehlberg:1968}
Fehlberg E.,  1968, NASA TR, R-287, 1

\bibitem[\protect\citeauthoryear{{Ferraz-Mello}}{{Ferraz-Mello}}{1999}]{Ferraz-Mello:1999}
{Ferraz-Mello} S.,  1999, \mn@doi [Celestial Mechanics and Dynamical Astronomy]
  {10.1023/A:1008370224264}, \href
  {https://ui.adsabs.harvard.edu/abs/1999CeMDA..73...25F} {73, 25}

\bibitem[\protect\citeauthoryear{{Hadden} \& {Lithwick}}{{Hadden} \&
  {Lithwick}}{2018}]{Hadden:2018}
{Hadden} S.,  {Lithwick} Y.,  2018, \mn@doi [\aj] {10.3847/1538-3881/aad32c},
  \href {https://ui.adsabs.harvard.edu/abs/2018AJ....156...95H} {156, 95}

\bibitem[\protect\citeauthoryear{{H{\'e}non}}{{H{\'e}non}}{1966}]{Henon:1966}
{H{\'e}non} M.,  1966, in {Kontopoulos} G.~I.,  ed.,  IAU Symposium Vol. 25,
  The Theory of Orbits in the Solar System and in Stellar Systems. p.~157

\bibitem[\protect\citeauthoryear{{Henrard} \& {Lemaitre}}{{Henrard} \&
  {Lemaitre}}{1983}]{Henrard:1983}
{Henrard} J.,  {Lemaitre} A.,  1983, \mn@doi [Celestial Mechanics]
  {10.1007/BF01234306}, \href
  {http://adsabs.harvard.edu/abs/1983CeMec..30..197H} {30, 197}

\bibitem[\protect\citeauthoryear{{Henrard}, {Lemaitre}, {Milani}  \&
  {Murray}}{{Henrard} et~al.}{1986}]{Henrard:1986}
{Henrard} J.,  {Lemaitre} A.,  {Milani} A.,   {Murray} C.~D.,  1986, \mn@doi
  [Celestial Mechanics] {10.1007/BF01238924}, \href
  {https://ui.adsabs.harvard.edu/abs/1986CeMec..38..335H} {38, 335}

\bibitem[\protect\citeauthoryear{{Lan} \& {Malhotra}}{{Lan} \&
  {Malhotra}}{2019}]{Lan:2019}
{Lan} L.,  {Malhotra} R.,  2019, \mn@doi [Celestial Mechanics and Dynamical
  Astronomy] {10.1007/s10569-019-9917-1}, \href
  {https://ui.adsabs.harvard.edu/abs/2019CeMDA.131...39L} {131, 39}

\bibitem[\protect\citeauthoryear{{Lecar}, {Franklin}, {Holman}  \&
  {Murray}}{{Lecar} et~al.}{2001}]{Lecar:2001}
{Lecar} M.,  {Franklin} F.~A.,  {Holman} M.~J.,   {Murray} N.~J.,  2001,
  \mn@doi [\araa] {10.1146/annurev.astro.39.1.581}, \href
  {http://adsabs.harvard.edu/abs/2001ARA%26A..39..581L} {39, 581}

\bibitem[\protect\citeauthoryear{{Mardling}}{{Mardling}}{2013}]{Mardling:2013}
{Mardling} R.~A.,  2013, \mn@doi [\mnras] {10.1093/mnras/stt1438}, \href
  {https://ui.adsabs.harvard.edu/abs/2013MNRAS.435.2187M} {435, 2187}

\bibitem[\protect\citeauthoryear{{Morbidelli}}{{Morbidelli}}{1996}]{Morbidelli:1996}
{Morbidelli} A.,  1996, \mn@doi [\aj] {10.1086/117979}, \href
  {https://ui.adsabs.harvard.edu/abs/1996AJ....111.2453M} {111, 2453}

\bibitem[\protect\citeauthoryear{Morbidelli}{Morbidelli}{2002}]{Morbidelli:2002Book}
Morbidelli A.,  2002, Modern celestial mechanics: aspects of solar system
  dynamics.
Taylor \& Francis London

\bibitem[\protect\citeauthoryear{{Murray}}{{Murray}}{1986}]{Murray:1986}
{Murray} C.~D.,  1986, \mn@doi [\icarus] {10.1016/0019-1035(86)90064-3}, \href
  {https://ui.adsabs.harvard.edu/abs/1986Icar...65...70M} {65, 70}

\bibitem[\protect\citeauthoryear{Murray \& Dermott}{Murray \&
  Dermott}{1999}]{Murray:1999SSD}
Murray C.~D.,  Dermott S.~F.,  1999, Solar system dynamics, 1 edn.
Cambridge University Press, New York, New York

\bibitem[\protect\citeauthoryear{{Namouni} \& {Morais}}{{Namouni} \&
  {Morais}}{2018}]{Namouni:2018}
{Namouni} F.,  {Morais} M.~H.~M.,  2018, \mn@doi [\mnras]
  {10.1093/mnras/stx2636}, \href
  {https://ui.adsabs.harvard.edu/abs/2018MNRAS.474..157N} {474, 157}

\bibitem[\protect\citeauthoryear{{Wang} \& {Malhotra}}{{Wang} \&
  {Malhotra}}{2017}]{Wang:2017}
{Wang} X.,  {Malhotra} R.,  2017, \mn@doi [\aj] {10.3847/1538-3881/aa762b},
  \href {http://adsabs.harvard.edu/abs/2017AJ....154...20W} {154, 20}

\bibitem[\protect\citeauthoryear{{Winter} \& {Murray}}{{Winter} \&
  {Murray}}{1997}]{Winter:1997}
{Winter} O.~C.,  {Murray} C.~D.,  1997, \aap, \href
  {http://adsabs.harvard.edu/abs/1997A%26A...319..290W} {319, 290}

\bibitem[\protect\citeauthoryear{{Wisdom}}{{Wisdom}}{1980}]{Wisdom:1980}
{Wisdom} J.,  1980, \mn@doi [\aj] {10.1086/112778}, \href
  {http://adsabs.harvard.edu/abs/1980AJ.....85.1122W} {85, 1122}

\makeatother
\end{thebibliography}


\bsp	
\label{lastpage}
\end{document}